\documentclass[lettersize,journal]{IEEEtran}

\usepackage{amsmath,amsfonts}
\usepackage{algorithmic}
\usepackage{algorithm}
\usepackage{array}
\usepackage{subcaption}
\usepackage{textcomp}
\usepackage{stfloats}
\usepackage{url}
\usepackage{verbatim}
\usepackage{graphicx}
\usepackage{cite}
\usepackage{xspace}
\usepackage{hyperref}
\usepackage{multirow}
\usepackage[numbers]{natbib}

\newcommand{\name}{VulMatch\xspace}

\newcommand{\mypara}[1]{\vspace{2pt}\noindent\textbf{{#1: }}}

\begin{document}

\title{\name: Binary-level Vulnerability Detection Through Signature}








\author{
    \IEEEauthorblockN{Zian Liu\IEEEauthorrefmark{1}\IEEEauthorrefmark{4}, Lei Pan\IEEEauthorrefmark{2}, Chao Chen\IEEEauthorrefmark{3}, Ejaz Ahmed\IEEEauthorrefmark{4}, Shigang Liu\IEEEauthorrefmark{1}, Jun Zhang\IEEEauthorrefmark{1}, Dongxi Liu\IEEEauthorrefmark{4}}\\
    \IEEEauthorblockA{\IEEEauthorrefmark{1}Swinburne University of Technology
    \\\{zianliu, shigangliu, junzhang\}@swin.edu.com}\\
    \IEEEauthorblockA{\IEEEauthorrefmark{2}Deakin University
    \\\{l.pan\}@deakin.edu.au}\\
    \IEEEauthorblockA{\IEEEauthorrefmark{3}Royal Melbourne Institute of Technology
    \\\{chao.chen\}@rmit.edu.au}\\
    \IEEEauthorblockA{\IEEEauthorrefmark{4}Data61, CSIRO
    \\\{ejaz.ahmed, dongxi.liu\}@data61.csiro.au}
}

\maketitle

\begin{abstract}
Similar vulnerability repeats in real-world software products because of code reuse, especially in wildly reused third-party code and libraries. 
Detecting repeating vulnerabilities like 1-day and N-day vulnerabilities is an important cyber security task. 
Unfortunately, the state-of-the-art methods suffer from poor performance because they detect patch existence instead of vulnerability existence and infer the vulnerability signature directly from binary code. 
In this paper, we propose \name to extract precise vulnerability-related binary instructions to generate the vulnerability-related signature.
\name detects vulnerability existence based on binary signatures.
Unlike previous approaches, \name accurately locates vulnerability-related instructions by utilizing source and binary codes. 
Our experiments were conducted using over 1000 vulnerable instances across seven open-source projects.  
\name significantly outperformed the baseline tools Asm2vec and Palmtree. 
Besides the performance advantages over the baseline tools, \name offers a better feature by providing explainable reasons during vulnerability detection. 
Our empirical studies demonstrate that \name detects fine-grained vulnerability that the state-of-the-art tools struggle with. Our experiment on commercial firmware demonstrates \name is able to find vulnerabilities in real-world scenario.
\end{abstract}


\section{Introduction}
\label{sec:intro}

Finding vulnerabilities or bugs in software is vital to improve its quality. 
Vulnerabilities and bugs tend to inherit in new software products due to the sluggishness of making up-to-date patches. 
A vulnerability detection paradigm is learning from existing vulnerable codes to find similar vulnerabilities and bugs. 
Human security analysts can learn from hundreds or thousands of existing vulnerabilities to gain experience and improve security awareness to manually find vulnerabilities or bugs by reviewing the code \cite{heffley2004can}. 
However, with an extremely large number of codes to review, there is an urgent call for automated methods to identify vulnerability codes directly or filter out potentially vulnerable codes for human experts to review later. 
Moreover, automatic vulnerability detection is in high demand due to replicated vulnerabilities spread by code reuse as a common practice in the software industry \cite{code_reuse,haefliger2008code}. 
Detecting the 1-day or N-day vulnerabilities in binary code is vital because of the unavailability of source code in many real-world scenarios.
This paper's research question is \textit{how to effectively and efficiently find similar vulnerabilities or bugs from existing ones}.

Automated detection methods have great advantages over manual analysis because binary code is notoriously difficult for humans to read and understand.  
Mainstream research consists of three genres: binary code similarity detection, patch existence detection, and vulnerability signature detection. 
\begin{itemize}
 \item \textbf{Binary Code Similarity Detection.} 
 Given a set of query binary samples, code similarity detection tools \cite{spain, BINCLONE, SMIT, Kam1n0, MBC, IDEA, Expose, binsequence, tracy, exediff, genius, binslayer, cxz2014, rendezvous, BEAGLE, fossil, SIGMA, COP, gemini, VULSEEKER, qbindiff, safe, innereye, aDiff, asm2vec, multimh, binhash, ks2017, bingo, IMF-SIM, binhunt, esh, GITZ, binjuice, tedem, xmatch} identify the best matching code snippets stored in the database with known vulnerable binary codes. 
 Code similarity-based vulnerability detection finds vulnerable binary code but introduces excessive false positives because patched binary codes usually have high similarity scores. 
 Furthermore, the similarity-based method is coarse-grained. They only output similar binary snippets at a large scale (e.g., function level). Since the function level binary code is usually large in scale and the vulnerability commonly only relates to several instructions, they can not explain specifically what instructions indicate the vulnerability.

 \item\textbf{Patch Existence Detection.} This genre of work \cite{fiber, pdiff, spain, patchscope} determines whether a patch exists in a query binary function. 
 This genre of work extracts patch code signatures and detects the existence of patch signatures in the query function. 
 However, it usually targets kernel binaries with debugging symbols like function names that are used to filter the query function and detect patch signatures. 
 Moreover, this genre of work fails to address the existence of vulnerability because the lack of patches does not equal the vulnerability's existence. 
 In the National Vulnerability Database (NVD), some Common Vulnerabilities and Exposures (CVEs) are vulnerable from some versions in a series, suggesting that the versions before the consecutive vulnerable versions do not contain patched code. 
 For instance, a project contains ten versions, but the versions between the third and the sixth are vulnerable, so its first two versions are not considered vulnerable because of the absence of patches.  

 \item\textbf{Vulnerability Signature Detection.} This genre of work \cite{vmpbl, viva, binxray} detects fine-grained vulnerability-related signatures in the binary code. 
 Existing works extract different instructions between two binary reference versions (i.e., a vulnerable version and a patched version). 
 Then they normalize the instructions and generate traces (i.e., blocks of normalized instructions) to form the vulnerability or patch signature. 
 However, extracting signatures directly at the binary level could introduce instructions irrelevant to vulnerabilities because the compiler replaces instructions with the same semantic and inlines functions. 
 For example, the source code line \texttt{bool fromfile=FALSE;} could be compiled to \texttt{mov [rsp+48h+var\_39], 0} in one version and \texttt{xor r14d, r14d} in other versions even using the same optimization flags (options). 
 The same variable is stored on the stack in the former version and the register \texttt{r14d} in the latter version. 
 If two versions inline another non-vulnerable function, and the inlined function has changed in the patched version, directly diffing the binary codes will include the changed instruction in the inlined function. 
 We reproduced the methods to understand such inaccurate cases and manually analyzed the corresponding output binary signatures. 
 We found that their methods introduce approximately 40\% vulnerability-irrelevant instructions into the signatures.
\end{itemize}

We propose a novel approach to generate accurate and fine-grained vulnerability-related signatures to address those research gaps. 
Firstly, we spend significant manual efforts pre-processing the data to include all the CVEs' information, each vulnerable function, the source code file it lies in, the affected versions, and the corresponding source code versions. 
To generate accurate and fine-grained binary code signatures, we generate source-code-level signatures and align them to binary-level signatures with the help of debugging information. 
Unlike existing work \cite{vmpbl,viva, binxray} that directly diff different binary versions to extract vulnerability signatures, we utilize source code to guide us to locate vulnerable binary code more accurately. 
Hence, we exclude many vulnerability-irrelevant binary code contents. 
To utilize non-trivial source-code information, \name processes the source code to prevent vulnerable functions from being inlined.
\name locates binaries from source code by handling different situations as described in \autoref{sec:Vulnerability and patch signature generation}. 
Note that the source code is only required for generating the signature and not for matching a given binary.
\name aims to find vulnerabilities in the query binary, which should not contain any debugging information and source code. 
We combine the information of source code, binary code, and debugging information to generate (learn) the signature accurately.
To match the binary-level signatures, we propose three signature types (i.e., add, delete, and change). 
To match the existence of fine-grained binary signatures rather than the whole-function-level similarity as the similarity-based genre, we create the binary signature with local control-flow information. 
The local control-flow information refers to the context instructions. It enriches the signature with unique features.  
To assist humans in understanding the decision made by \name, a user interface interpreting the matched binary signatures shows the matched signatures and the match score in the binary.

To evaluate the utility of \name, we prepared seven popular open-source projects with well-documented vulnerability information. 
In total, there are 906 CVEs, including 1281 vulnerable functions.
Our results demonstrate that \name outperforms two state-of-the-art vulnerability detecting tools --- Asm2vec and Palmtree by approximately 9\% and 6\% more top-1 score, 80\% and 79\% less mismatch score, respectively.
We also demonstrate how \name assists humans in understanding its detection results in terms of interpretability. We experiment with commercial firmware to demonstrate \name is practical to find real-world vulnerabilities.
We perform in-depth research on the vulnerability and signature types and their distribution in the dataset.

This paper makes the following contributions:
\begin{itemize}
  \item We propose a novel approach to extract, store, and match the vulnerability-related signatures. 
  We have implemented the approach into a tool called \name that is open-source and publicly accessible on GitHub \footnote{{The source code is available at https://github.com/Vulmatch/Vulmatch.git
}}.
  \item To facilitate the human to understand \name's results and the reason \name decides whether the query binary contains vulnerability or not, we provide interpretability functionality in \name.
  \item We perform in-depth analysis on vulnerability and signature types and their distribution across all datasets. 
  We inspect each dataset's top three vulnerability types and different signature types with the average signature size.
\end{itemize}


\section{A Motivating Example}
\label{sec:mot}

\mypara{Terms definition} \textbf{Block} refers to a set of consecutive binary (assembly) instructions split by the control-flow-related instructions (e.g., jump instructions). 
An assembly function consists of various blocks connected to each other. 
Blocks are connected together to represent the assembly code's control-flow graph (CFG), as shown in \autoref{fig:mot}. 
Note that in this paper, we will use the terms `binary code' and `assembly code' interchangeably. 
\textbf{CVE} refers to the vulnerability in the function. 
One CVE may correspond to multiple vulnerability-related instructions and multiple signatures.

\begin{figure}[!t]
\centering
\includegraphics[width=0.47\textwidth]{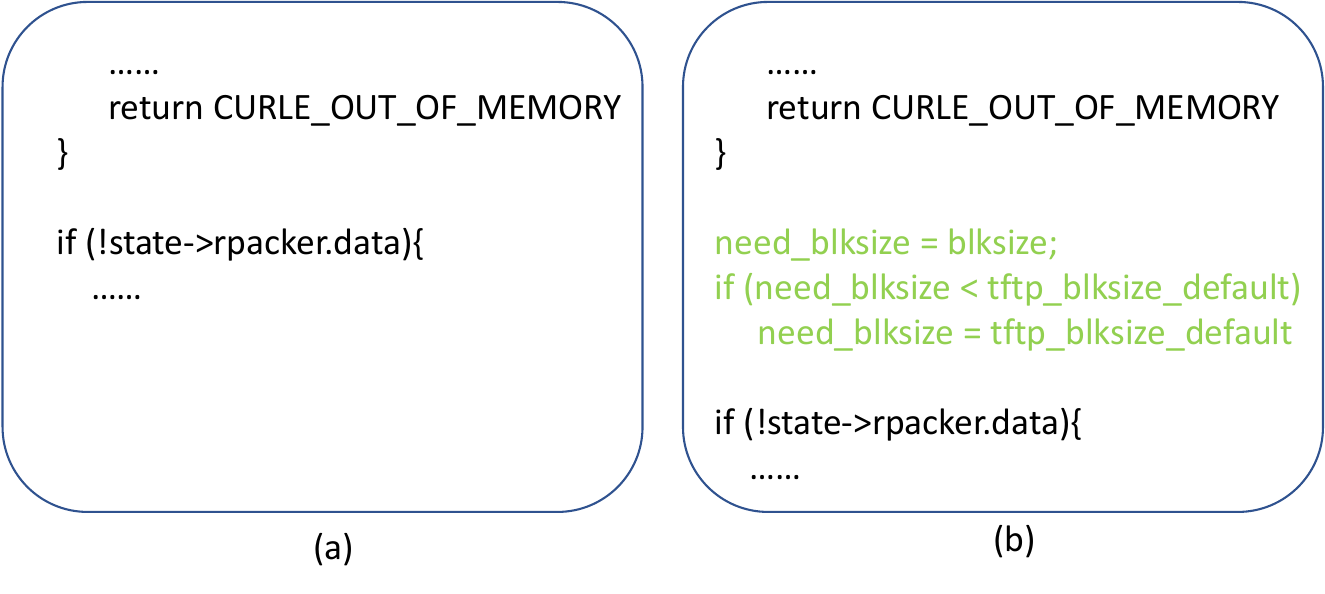}
\caption{An example vulnerable function \texttt{tftp\_connect} selected from CVE-2019-5482. (a) lists pre-patch source code, and (b) lists post-patch source code. Green lines are the patched source lines. Other lines remain intact across the two versions.}
\label{fig:mot1}
\end{figure}

\begin{figure}[!ht]
\centering
\includegraphics[width=0.35\textwidth]{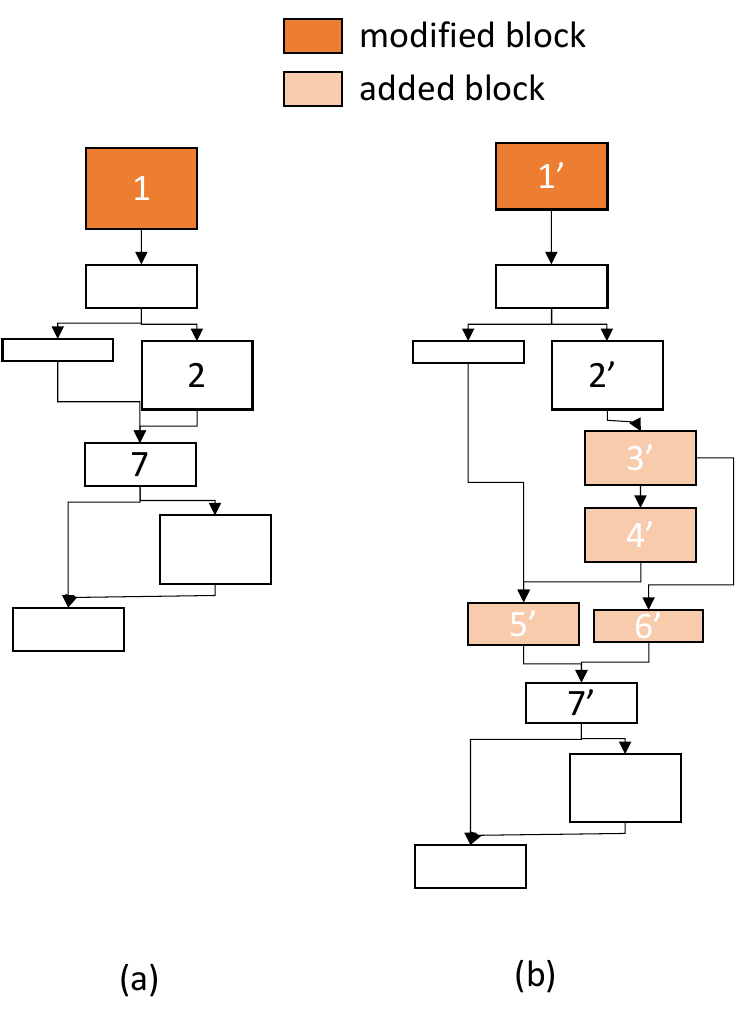}
\caption{Corresponding binary code CFG of function \texttt{tftp\_connect} presented in \autoref{fig:mot1}. (a) refers to pre-patch version, and (b) refers to post-patch version. Block 1' is a modified block and blocks 3', 4', 5', and 6' are added blocks. Other blocks remain intact.}
\label{fig:mot}
\end{figure}

\autoref{fig:mot1} shows the source code snippets of the vulnerable function \texttt{tftp\_connect} from CVE-2019-5482 before and after the patch, where green lines are the patched instructions. 
\autoref{fig:mot} shows the corresponding binary code structure.
\autoref{fig:mot}(a) is the vulnerable version (before patch), and \autoref{fig:mot}(b) is the patched version. 
The binary code samples were built from the source code snippets using an identical compilation configuration with additional debugging information. 
Since the patch in the example comprises two kinds of changes through added and modified instructions, they are listed using different colors. 
Specifically, block 1' is a modified block of instructions, and blocks 3', 4', 5', and 6' are added blocks of instructions. 
Other blocks remain intact.

Similarity-based lines of work compare the whole functions' similarities before identifying a potentially vulnerable function if the function is similar to the vulnerable function. 
They focus on the whole function similarity rather than vulnerability-related instructions, resulting in poor granularity. 
Furthermore, they fail to distinguish the vulnerable and the patched functions since they are regarded as similar. 
Patch-detection lines of work first use the similarity lines of work to filter potential similar functions. 
They assume to select a similar function by name to detect the existence of the patch, where the binaries are Linux kernel binaries. 
However, as mentioned in \autoref{sec:intro}, if the patch does not exist, it does not necessarily mean that the binary is vulnerable. 
Existing binary signature-based methods directly diff the vulnerable and the patched binary versions and assume the different binary instructions are all vulnerability-relevant. 
However, we reproduced their methods with a manual analysis of the results and found that up to 40\% vulnerability-irrelevant instructions were included.

After manually inspecting the source code snippets and the corresponding binary samples, we found that only blocks 3' and 6' are the actual patched blocks corresponding to green lines in \autoref{fig:mot1}. 
Other changed blocks (i.e., blocks 1', 4', and 5') are not aligned with any changed source lines, but they map to the unchanged source code lines.
The changes in blocks 1', 4', and 5' were due to replacing instructions with the same semantics, which is the indirect impact of the patched instructions. 
Existing work in \cite{vmpbl, binxray, viva} failed to identify these blocks as unchanged code.
To rectify this issue, \name generates and matches the vulnerable signature with the guidance of the source code. 
We introduce the three steps of \name as follows:

\mypara{Step1: Locating Signature Instructions.}
We use the \texttt{diff} tool to measure source-code-level differences. 
Diff can detect and output a list of changed sites, added sites, and deleted sites. 
In the example shown in \autoref{fig:mot}, diff scans the source code in \autoref{fig:mot1} and outputs one added site. 
Subsequently, we use the debugging information in the binary code (i.e., the source-binary lines mapping) to locate the patched binary lines. 
In the diff's output, the changed site contains the source code lines in both pre-patching and post-patching versions. 
However, the diff output for add site only contains the added source lines in the post-patching version (e.g., green lines in \autoref{fig:mot1} (b)). 
Since the added instructions do not exist in the pre-patching version, diff has no outputs for the pre-patching version.
Therefore, for add type signature, an additional process will take place later in step 2 to find vulnerability-related instructions in the pre-patch version. 

\begin{figure*}[!t]
\centering
\includegraphics[width=\textwidth]{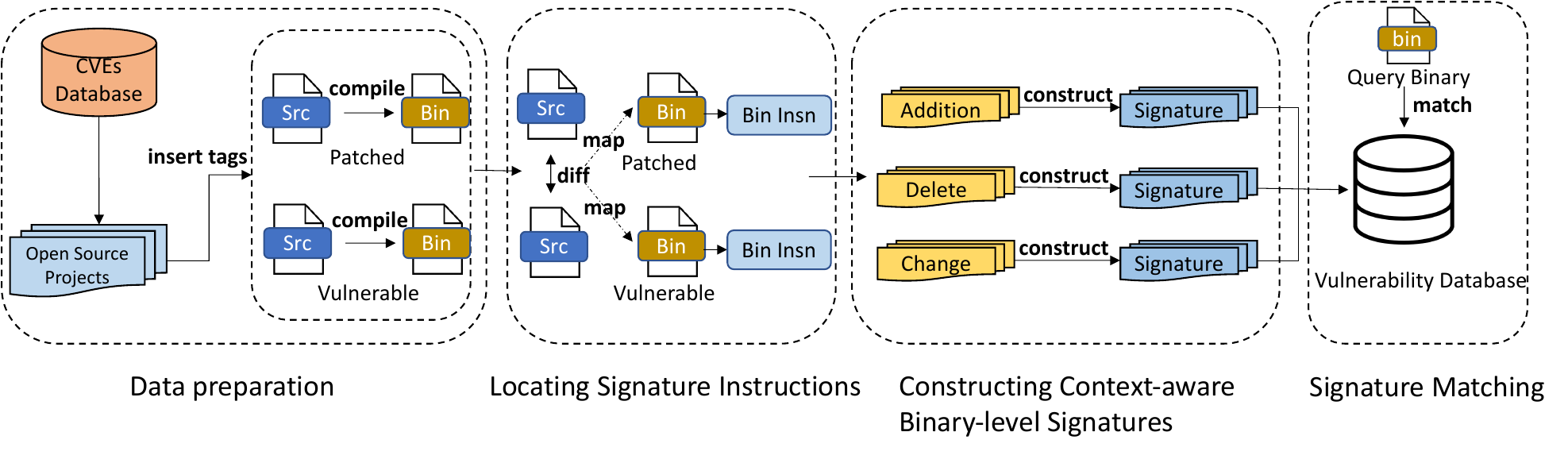}
\caption{\name consists of four steps: Data Preparation, Locating Signature
Instructions, Constructing Context-aware
Binary-level Signatures, and Signature Matching. \textit{Src} is short for source code. \textit{Bin} is short for binary code. \textit{Insn} is short for instruction.} 
\label{fig:methodology}
\end{figure*}

\mypara{Step2: Constructing Binary-level Signatures.} 
In this step, we use the located binary instructions in step 1 to construct the binary signatures. We store both vulnerable and patched signatures in the database.
For the added signature, we still need to generate its vulnerable binary signature even if diff outputs nothing at the source-code level. 
We cannot directly consider the absence of the added (patched) instructions to imply vulnerability because another random function does not necessarily have the added (patched) instructions. 
The random function needs to be not vulnerable. 

Therefore, we need to use the vulnerability-related instructions in \autoref{fig:mot}(a) to construct a vulnerability signature. 
Note that the added instructions are inserted between block 2 and block 7. 
Therefore, in the vulnerable version \autoref{fig:mot}(a), the control flow from block 2 to block 7 implies the vulnerability. 
In the patched version \autoref{fig:mot}(b), the control flow from block 2' to block 3' and from block 3' to block 6' implies patch existence.  
Therefore, we store the control flow from block 2 to block 7 as the vulnerable signature. 
Additionally, we store the control flow from block 2' to block 3' as the patch signature.

\mypara{Step3: Matching Signatures.}
For a query (unknown) binary, we check whether the vulnerability is related to CVE-2019-5482 stored in the database. 
If we store multiple signatures in the database for one CVE, we will check each signature and aggregate an overall score. 
For the changed or deleted signatures, we detect the percentage of the matched vulnerability instructions with respect to the query binary.
For example, if the changed signature block contains 5 instructions and 3 of them exist in some block in the query function, then the score of the changed signature is 3/5=0.6. 
For the added signature, we check the existence of the control flow (e.g., the stored control flow from block 2 to block 7 in \autoref{fig:mot}(a)). 
We count how many matched instructions exist in the query function for each control flow. 
If there are 10 instructions in blocks 2 and 7, and we found a similar control flow in the query function with 8 instructions matched, then the score is 8/10=0.8. 
However, if we detect the existence of the patch signature in the query function, we directly consider the query function contains a patch and output that signature score as 0. 
Finally, we average all the signature scores according to their weights (instruction sizes) to derive the overall score.

To summarise, the input of our proposed method to produce the vulnerable binary signatures are: 1) CVE information, including the last vulnerable version, first patched version, and vulnerable function name. 2) Source code with different versions. 
Then, in the query phase, the input could be an unknown binary code without debugging information and source code. 
The output is a list of potentially matched CVEs with the similarity score. 
Compared to existing methods, \name yields more accurate binary signatures with less vulnerability-irrelevant instructions. \name is able to accurately predict the vulnerable sites in the query binary rather than only giving a similar code. \name is able to accurately match the real vulnerable binary code with fewer false positives among several similar binary code snippets.

\section{Methodology}
\label{sec:methdlg}


This section presents the design of \name. \name's four components are shown as \autoref{fig:methodology}.

\subsection{Data Preparation}
\label{sec:data preparation}
We collect many already well-studied vulnerabilities from several publicly-available open source projects to build the vulnerability database. 
According to \cite{viva}, vulnerabilities tend to be fixed in new versions of software releases. 
Thus, the vulnerability-related versions consist of the last pre-patching and the first post-patching versions.
The last pre-patching and the first post-patching version will be used later to extract the signatures.
We download all the vulnerability-related versions for each project and record each CVE's information. 
Specifically, for each CVE, we record its related vulnerable source code file name and the vulnerability-related functions within them. 
We also record each CVE's affect versions for later preparing testing binaries for evaluation.

\mypara{Challenges} 
Not all vulnerability-related functions exist in the compiled binary code due to the automatic function-inlining behavior. 
Automatic function-inlining refers to merging a function \texttt{FuncA} into another function \texttt{FuncB} that calls back \texttt{FuncA}. 
If vulnerable functions are inlined, it would be challenging to locate them in the binary code. 
This case holds even if we manually turn off the function-inline option during compilation. 
Hence, it is challenging for us to generate binary signatures.

\mypara{Solution} 
We need to ensure that the database contains no inlined functions in the compiled binaries.
\name automatically analyzes the source code files and edits the functions in the source code files to inform the compiler not to inline the function.
Technically, \name inserts a non-inline tag \texttt{\_\_attribute\_\_((noinline))} before each vulnerable function in all related versions to preserve the tagged functions in the compiled binary code. 
For each CVE, \name loads the CVE's information to retrieve its vulnerable source code files along with the corresponding vulnerable functions. 
Then for each related version (i.e., the last pre-patching version and the first post-patching), \name analyzes the vulnerable source code file to locate the vulnerable functions and automatically insert no-inline tags. 
Finally, we compile these versions into binaries with the same default compilation options.

\subsection{Locating Signature Instructions and Challenges}
\label{sec:Vulnerability and patch signature generation}
We generate signatures related to vulnerabilities and patches using the source codes and compiled binary codes. 
For each vulnerable function, we generate its signatures in two steps --- 1) generate source-level vulnerability-related instructions, 2) locate vulnerability-related binary instructions through mapping.

\subsubsection{Generating Source-level Vulnerability-related Instructions} 
We prepare the last pre-patching and the first post-patching versions using the information we retrieved in \autoref{sec:data preparation}. 
Subsequently, we generate vulnerability and patch-related signatures on the source code level. 
We use the \texttt{diff} tool\footnote{https://man7.org/linux/man-pages/man1/diff.1.html} to extract source-code-level patched instructions.
There are three types of source-code-level patches in the \texttt{diff} outputs. 
1) Added instructions that are used in the patched version and absent in the vulnerable version, as shown in \autoref{fig:add_example}.
2) Deleted instructions that are removed from the vulnerable version and absent in the patched version as shown in \autoref{fig:delete_example}. 
3) Changed instructions that are updated from the vulnerable version to the patched version, as shown in \autoref{fig:change_example}. 
The changed instructions usually share the same context instructions among the two versions.

\subsubsection{Locating Vulnerability-related Binary Instructions through Mapping} 
We use the source-to-binary mapping with the binary's debugging information to locate the source code's corresponding assembly instructions.
Although \name employs the simple idea, there are practical challenges primarily in two aspects. 
\begin{enumerate}
 \item Asymmetric source-binary mapping: it is challenging to map source line changes in the source code files (e.g., .cpp or .c file) to the corresponding binary file,
 \item Identification of vulnerability-specific source lines.
\end{enumerate}


\begin{figure}[!t]
\centering
\includegraphics[width=0.37\textwidth]{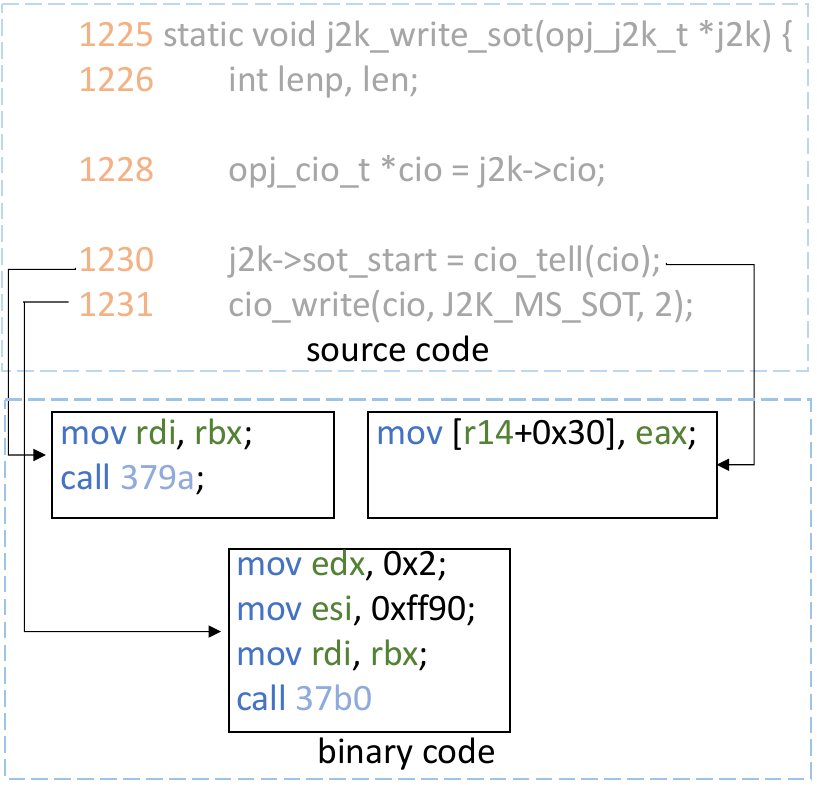}
\caption{An example of a missing match between source code and binary code. The first two lines 1226 and 1228 do not have any mapping instructions in binary code because the assembly code does not need to specify the type information for functions and variables. Line 1230 maps to two different basic blocks. Line 1231 maps to one basic block. This example is extracted from openjpeg version 1.5.0.}
\label{fig:missingmap}
\end{figure}

\mypara{Two challenges to map source code files} 
\begin{itemize}
    \item \textbf{Challenge1: Asymmetric source-binary mapping.} 
    Not all the source code lines have a matching binary code instruction. 
    For example, \autoref{fig:missingmap} shows an example of missing mapping between source code and binary code. 
    Lines 1226 and 1228 declare new variables but do not map to any binary instructions because the variables at the binary level are directly used without explicit type declaration due to the binary code convention. 
    
    \item \textbf{Solution1:} 
    Generally, the source code lines declaring new variables (e.g., line 1226 and 1228 in \autoref{fig:missingmap}) do not have a mapping binary code because of binary code convention.  
    However, it does not affect finding the binary signatures. 
    We further elaborate on the following two cases: 
    1) If a new variable declaration is added, it must be used later in some other source code lines, implying that the correlated source lines still exist after diffing source codes of the patched and vulnerable functions. 
    2) If a variable's name is changed, the source code referring to that variable must change, which is detected by diffing the source codes. 
    For a variable with type change (e.g., change from a defined structure \texttt{structA} to an updated structure \texttt{structA'}), source code lines using that variable tend to change because of different type usage (e.g., defining different fields in the different structure type). 
\end{itemize}

\begin{figure}[!t]
\begin{subfigure}[b]{0.5\textwidth}
         \centering
         \includegraphics[width=\textwidth]{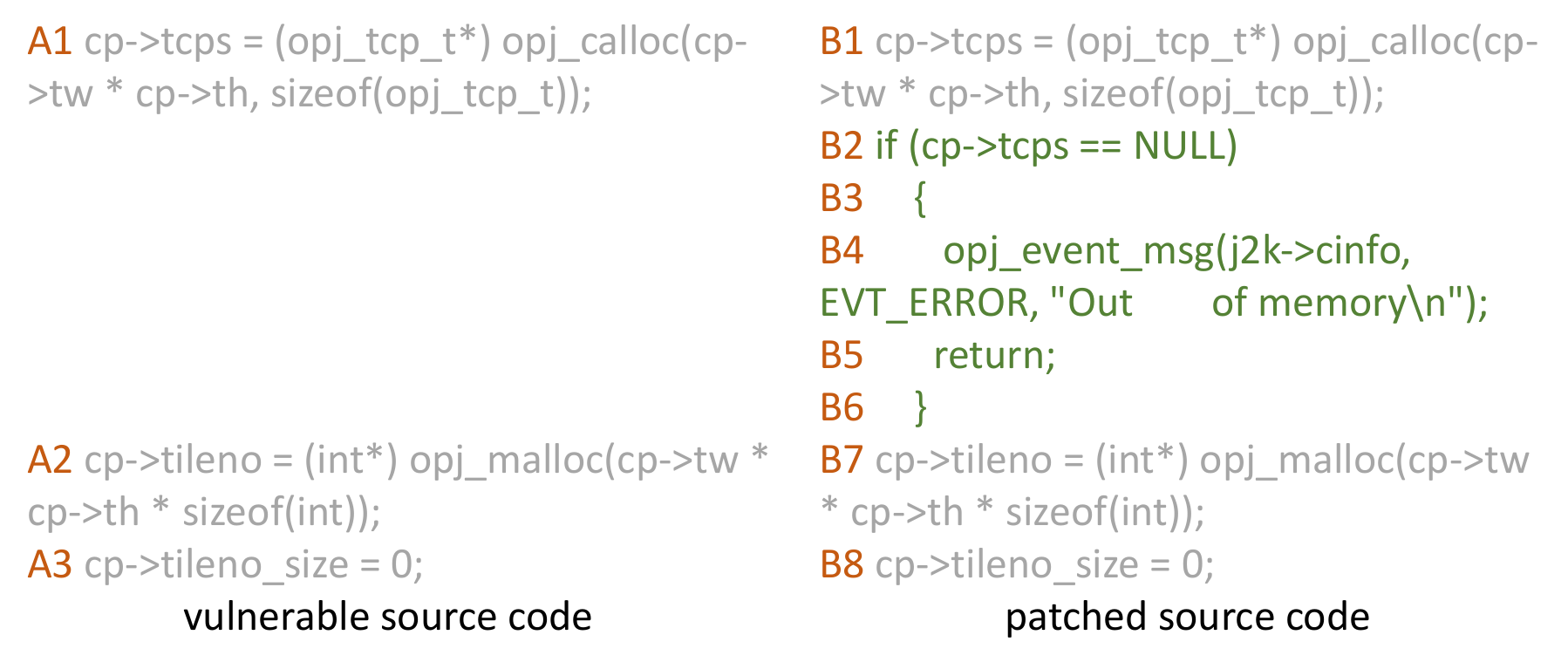}
         \caption{Add Type}
         \label{fig:add_example}
\end{subfigure}

\begin{subfigure}[b]{0.5\textwidth}
         \centering
         \includegraphics[width=\textwidth]{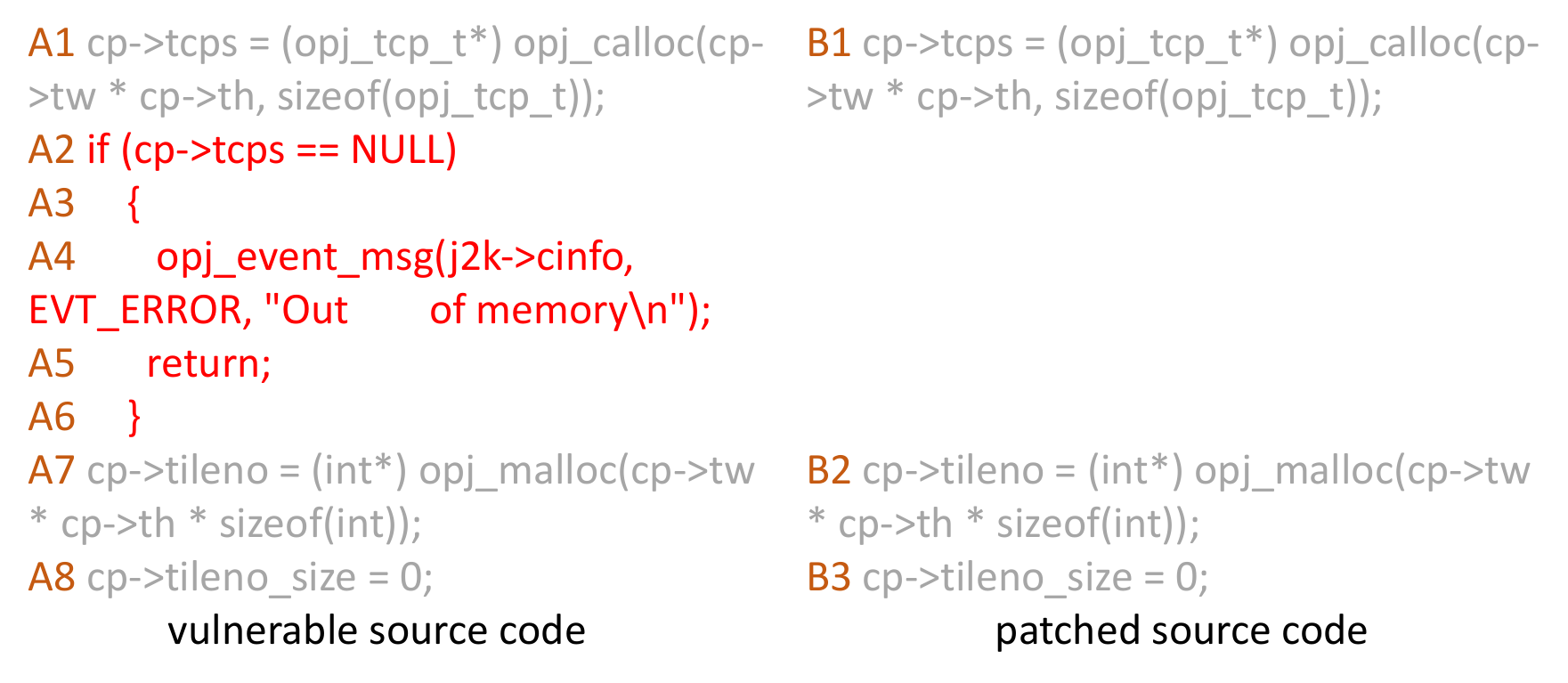}
         \caption{Delete Type}
         \label{fig:delete_example}
\end{subfigure}

\begin{subfigure}[b]{0.5\textwidth}
         \centering
         \includegraphics[width=\textwidth]{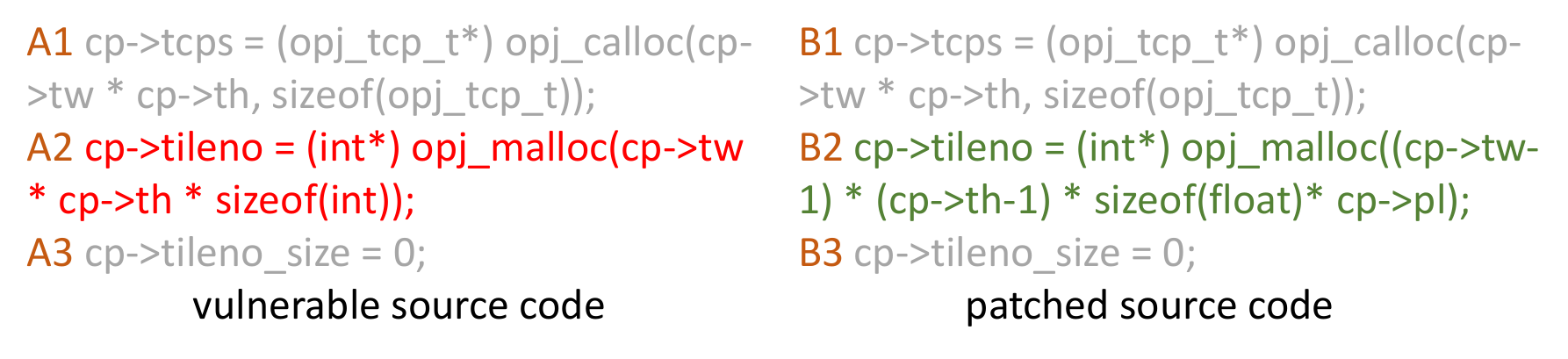}
         \caption{Change Type}
         \label{fig:change_example}
\end{subfigure}

\caption{Examples of add, delete and change types. Green lines are the newly added or changed instructions in the patched version. Red lines are the deleted or changed lines in the vulnerable version. Grey lines are the intact lines. }
\label{fig:addExample}
\end{figure}

\begin{itemize}
\item \textbf{Challenge2: Identification of vulnerability-specific source lines.} 
  The add type signature is challenging to represent. 
  Because the add type signature only exists in patched versions, the added instructions imply the existence of a patch rather than the vulnerability itself. 
  Therefore, there are no direct vulnerable instructions from the vulnerable version. 
  For example, \autoref{fig:add_example} shows an example of the add type signature in the source-code level. 
  Green lines (lines B2 to B6 on the right-hand side) are the added lines in the patched version, and grey lines are the unchanged lines across the two versions. 
  The absence of the green lines in the vulnerable version implies a vulnerability.
  However, other random functions may lack added instructions without the same vulnerability. 
  Therefore, the lack of added instructions cannot be directly used as the vulnerable signature. 
  We need to infer the vulnerability signature in the vulnerable version to detect vulnerability existence.
 
\item \textbf{Solution2:} 
To represent add type vulnerability signature, our solution is to focus on the context. 
For example, lines A1 and B1 in \autoref{fig:add_example} are unchanged in the two versions. 
A1 flows to A2 in the vulnerable version,  while B1 flows to B2 in the patched version. 
The control flow from the unchanged instruction A1 to the following instruction A2 is regarded as the vulnerability signature in \name. 
Conversely, the control flow from B1 to B2 is regarded as a patch signature. 
Since the added instructions are inserted at some point within the function, they must have identical context instructions (e.g., A1 and B1 in the example) with different subsequent instructions (e.g., A2 and B2). For simplicity, we explain this concept at the source code level. But we extract add type signatures at the binary level. For more details refer to \autoref{sec:form}.
\end{itemize}

\subsection{Constructing Context-aware Binary-level Signatures}
\label{sec:form}

We construct the binary-code-level signatures before storing them in the database for signature matching. 
Simply storing the sets of instructions in the database as vulnerable signatures and detecting those signatures' existence in the query binary code may not be beneficial. 
As mentioned in \autoref{sec:Vulnerability and patch signature generation}, added instructions in the patched binary cannot directly be used to form a vulnerability signature because it only indicates patches. 
The term \textit{context} refers to the adjacent blocks' instructions of the vulnerable binary instructions. 
The vulnerable binary instructions are usually short.
If we generate signatures by simply concatenating those instructions into a sequence, the signature may carry inadequate information to prevent false positives.
Therefore, we propose to form new structures by combining the context and the vulnerable instructions. 
Our newly combined structure gives the signature adequate uniqueness to boost the performance of signature matching.
We propose to build the context around the vulnerable signature instructions through generalization to reduce false positives. 
For instance, the extracted signature instructions size is small (e.g., only 3 instructions).
Checking the existence of signature instructions without context information makes the signature not unique enough, leading to excessive mismatches (false positives).

Since the added instructions in the patched version have blocks directly preceding them, the counterpart preceding blocks in the vulnerable version should have different instructions following them. 
Therefore, we capture local control flows around the preceding blocks in the vulnerable version to represent the vulnerability signature.

\begin{figure}[!t]
\centering
\includegraphics[width=0.47\textwidth]{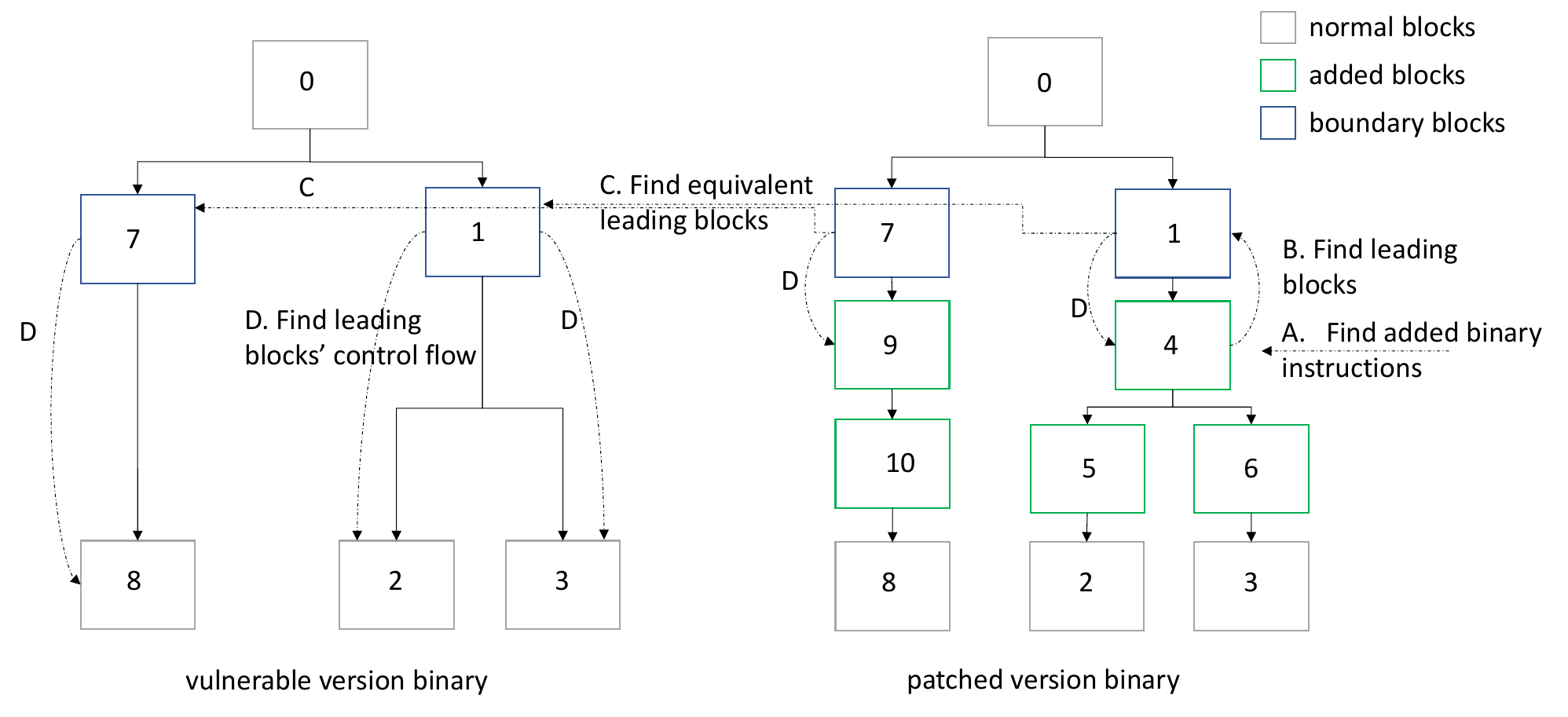}
\caption{An example of binary-code-level add signature and the steps to extract the corresponding binary signature.}
\label{fig:leading}
\end{figure}

We propose to generate the binary-level signature with control-flow information. 
Firstly, we define several terms. 
\begin{itemize}
   \item \textbf{Add Batch}. When newly added source code snippets are mapped to binary code blocks, the newly added blocks could either be directly connected to each other (e.g., block 4 and 5 in \autoref{fig:leading}) or separate from each other (e.g., block 4 and 9 in \autoref{fig:leading}).
   An \textbf{add batch} is made up of the added blocks that are strongly connected to each other.
   As shown in \autoref{fig:leading}, block $(4, 5, 6)$ and block $(9, 10)$ are two add batches.
   \item \textbf{Leading Blocks.} 
   The \textbf{leading block} is the unchanged block immediately preceding an added batch.
   As shown in \autoref{fig:leading}, blocks 1 and 7 are two leading basic blocks because they immediately precede two add batches.
   \item \textbf{Parents-children Structure.} We define a parents-children structure to store the control flow and literal information for add and change signatures. 
   Specifically, in one parents-children structure, we have an initial block from the function as the parent. 
   We include the chosen block's children blocks in the function into the parents-children structure. 
   Conversely, we can select a child block before including its parents to establish a parents-children structure.
   \item \textbf{Block List Structure}. We define a block list structure to store only the literal information when control-flow information is not available or unnecessary. 
   In one block list structure, we store all the vulnerable binary instructions grouped by blocks.
\end{itemize}

We store both vulnerable and patch signatures. 
Vulnerability signatures are generated from the instructions in the vulnerable version. 
This signature type consists of parents-children structures or block list structures. 
Patch signature consists of the instructions that only exist in the patched version and only consists of the block list structure. 
Patch signature is used to reduce the false positives further. 
Despite the vulnerability match score, the patch signature directly implies a patch. 
The vulnerable signatures contain three types: 1) add, 2) delete, and 3) change.
Those signatures have different structures to capture different information because different signature type has different nature. 
We capture various information for different signature types to enrich the signature information.

For the \textbf{add type signature}, to locate the add type binary signatures, we \textbf{A)} retrieve the added binary instructions in the patched version (i.e., the output of the operations in \autoref{sec:Vulnerability and patch signature generation}). 
\textbf{B)} We find the leading basic blocks in the patched version binary.
\textbf{C)} We find its counterpart leading basic block in the vulnerable version binary. 
\textbf{D)} We include the vulnerable binary's leading basic blocks' children blocks as a parents-children structure in the signature. 

For the \textbf{delete type signature}, we directly locate the mapping binary instructions and store those instructions into block list structures as delete signatures since the deleted instructions usually map to multiple blocks. Since the mapped blocks are usually sufficient in amount, lexical information already makes the signature unique for matching. If we record their control-flow information we will use excessive parents-children structures. 
We exclude any patch signature for this signature type because the patched version does not have any unique instruction that does not exist in the vulnerable version.

The \textbf{change type signature} has two categories, including one-block-change and many-blocks-change. 
Many-block-change means the changed instructions are distributed in multiple blocks (i.e., distributed in neighbor blocks or blocks that are not directly connected). 
One-block-change is the case if all the changed instructions are accommodated in one block in the binary code.

\begin{table*}[!t]
\centering
\caption{Information of the seven selected open-source projects.}
\label{tab:projects}
\footnotesize
\begin{tabular}{l|l|c|c|c|c|c|c|c}
\hline
\textbf{Project} & \textbf{Domain} &\textbf{Versions(\#)} & \textbf{Binary Files(\#)} & \textbf{.c Files(\#)} & \textbf{.h Files(\#)} & \textbf{CVEs(\#)} & \textbf{Vulnerable Functions(\#)} & \textbf{Avg Size}\\
 \hline
Tcpdump &Packet Analyzer& 20 &152 &167 &78 &192 &213 &20.45 \\

Curl &Data Transferring&  67 &315 &419 &197 &111 &231 &44 \\

OpenSSL &Protocols& 51 &755 &903 &243 &114 &220 &205 \\

Openjpeg &Image Processing&  15 &104 &205 &139 &94 &187 &24.50 \\

LibPNG &Image Processing& 63 &39 &36 &14 &52 &50 &6.90 \\

Libtiff &Image Processing& 30 &69 &102 &24 &142 &169 &12.30 \\

FFmpeg &Multimedia Processing& 104 &1206 &1591 &629 &201 &211 &584 \\

\hline
\textbf{Total} & various & 350 &2640 &3423 &1324 &906 &1281 &897.15 \\
\hline
\end{tabular}
\end{table*}

\mypara{Many-block-change}
If the change is many-block-change, we will need to record both control-flow and lexical information in the database since the change sites are usually small in size. This category of signature provides rich information as it contains adequate lexical information (i.e., binary instructions) from multiple blocks or control-flow information between those blocks. 
Therefore, for each block in a many-block-changes structure, if its neighbor (i.e., either predecessor or successor block) is a change block, we include this neighbor to form a parents-children structure. 
If none of its neighbor blocks is changed, all changed instructions are grouped as a block in the signature. Note that if the many-block-change contains a deeper level other than two levels (i.e., the level of parents-children structure), we use multiple parents-children structures to cover all the strongly connected blocks. For example, if block $A$ flows to block $B$, and block $B$ flows to block $C$, we will have two parents-children structures to cover the flow from $A$ to $B$ and from $B$ to $A$ respectively.

\mypara{One-block-change}
Conversely, if the change is a one-block-change, the information is limited because we only have lexical information without control flow information. 
Thus, we need to add more control flow information to enhance the signature and reduce potential mismatch. 
We include its parent blocks in a parents-children structure to enhance the signature. 
We include the children blocks in the parents-children structure if it has no parent block.  

\mypara{Patch signatures}
We generate signatures for patches. 
After we generate vulnerable signatures as above, we diff the vulnerability-related sites in both versions. 
We identify the instructions that only exist in the patched version and store it using a block list structure as the patch signature.

\subsection{Signature Matching}
We detect the vulnerability's existence by using both vulnerability and patched signatures. 
For the add signature, we search for each vulnerable parents-children structure in the query binary code. 
Then, we check for the existence of a patch signature. 
If a patch is found, the function is directly considered patched. 
For the delete signature, we search for the existence of the blocks from the block list structures. 
We do not match patch signatures for the delete type because the delete type does not has unique instructions in the patched version. 
For the change signature, we search for the existence of each parents-children structure or block list structure in the query binary code. Subsequently, we check the existence of the patch signature. 
If the patch is found through a query, the function is considered patched (denoted by $P=1$);  otherwise, $P=0$. 

We propose a measurement of the vulnerability existence score ($Sim$) to demonstrate the probability of the query function containing a given vulnerable signature. 
Specifically, a final score of vulnerability existence is calculated as follows:
\[Sim=\begin{cases}\frac{\Sigma_{i=1}^{len(S)}Matched(S[i])}{\Sigma_{i=1}^{len(S)}Total(S[i])} &\text{if }P=0\\0 &\text{if }P=1
\end{cases}\]
where $Sim$ represents the result similarity score to the vulnerable signature. 
$S$ represents one vulnerable signature. 
A signature consists of one or multiple structures (a structure is either parents-children structure or block list structure). 
$len()$ calculates the number of structures regarding an input signature.
$S[i]$ represents a structure. 
$Matched()$ calculates the number of instructions matched between the input structure and the given query binary function. 
If the structure is parents-children structure $PS$, $Matched()$ searches through the query binary to find the similar parents-children structure $PS'$ with the maximum similarity. 
Then $Matched()$ counts the instructions shared between $PS$ and $PS'$. 
If the structure is a block list structure, $Matched()$ finds all the blocks with the maximum similarity to each block in the block list structure before $Total()$ aggregates the total instruction number of the input structure.

\section{Evaluation}
\label{sec:eval}

\subsection{Experimental Setup}

\mypara{Data Collection}
We collected source code for seven open-source projects, including OpenSSL, OpenJPEG, FFmpeg, TCPDUMP, LibTIFF, cURL, and LibPNG. 
These projects are selected from diverse domains like communication protocols, image processing tools, and network traffic analyzers.
After manual analysis, we extracted 906 CVEs corresponding to 1,281 vulnerable functions.
\autoref{tab:projects} lists the versions, application domains, CVE information, vulnerability, and code-related information.

\mypara{Baseline Tool Selection and Testbed}
We prepared two state-of-the-art baseline tools Asm2Vec \cite{asm2vec} and PalmTree \cite{Palmtree}, because of their popularity and excellent performance in vulnerability detection.  
We ran \name and Asm2vec on an Intel NUC kit (NUC8i5BEH) with an i5-8259U processor and 16 GB memory. 
Since Palmtree is a deep learning-based approach and requires intensive GPU power, we ran it on an accelerator cluster of high-performance computer (HPC) systems with 456 NVidia Tesla P100, 114 Dual Xeon 14-core E5-2690, and 256 GB memory.

\mypara{Project Compilation}
As mentioned in \autoref{sec:data preparation}, we compile all the versions relating to each vulnerability (i.e., the last version before patching and the first version after patching) of the project to generate binary code instances. 
Depending on the project, we use the projects' default compiling flags, either \texttt{-O2} or \texttt{-O3}. 
For each project, we use identical compiling flags for building. 
So when we diff the compiled binary code to generate vulnerability and patch signatures, the compiling options are the same. This minimizes the differences in binary codes and is the common practice as \cite{binxray, viva} to help find vulnerable instructions. 
At compile time, we set the debugging symbol option to acquire source-binary instructions mapping that will serve as ground truth.

\begin{figure}[!t]
\centering
\includegraphics[width=0.5\textwidth]{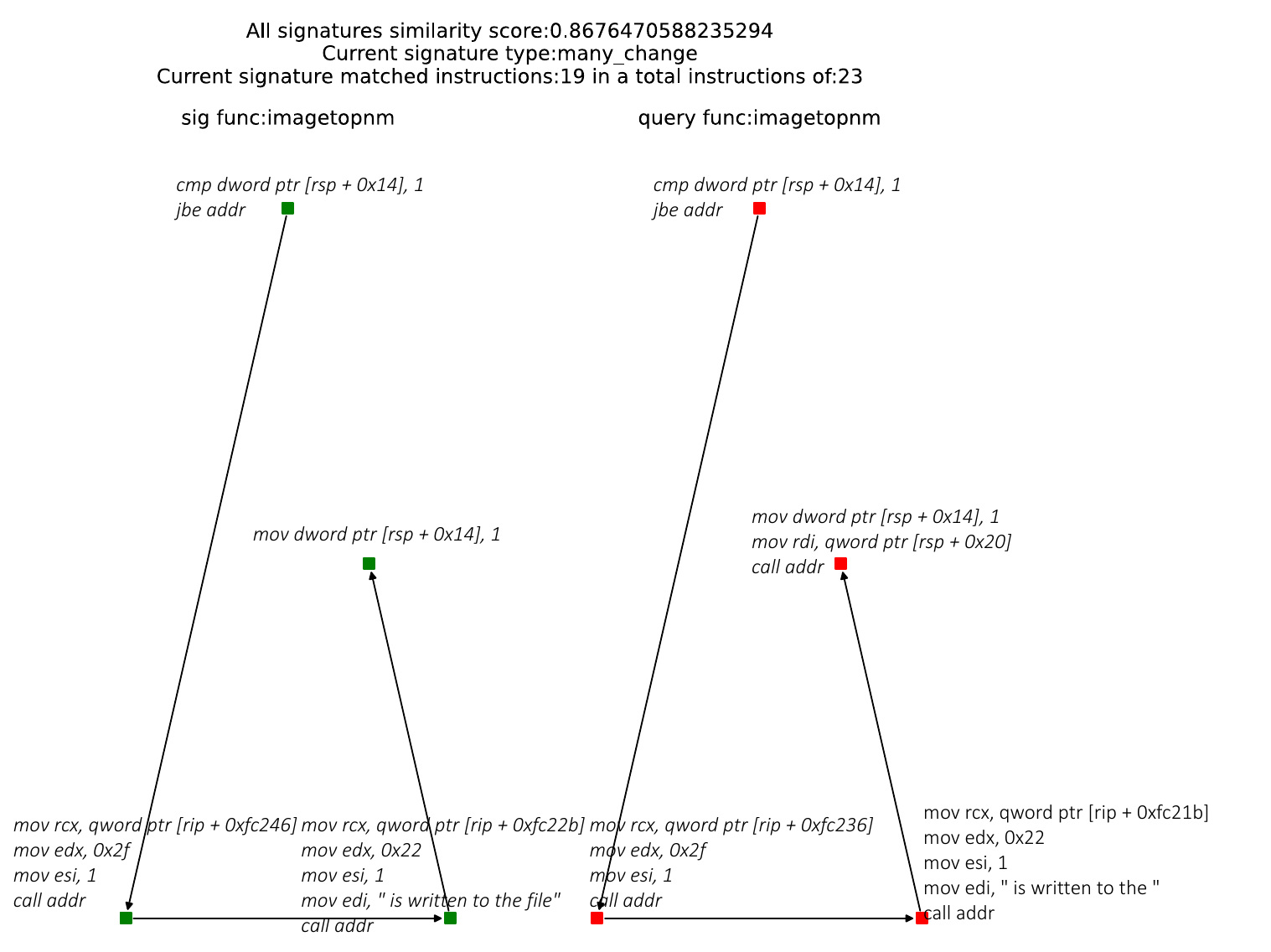}
\caption{Interpreting a \texttt{many-block-change signature} matching. The left-hand side is the generated vulnerable signature, and the right-hand side is the matched instructions in the query binary.}
\label{fig:sig3}
\end{figure}

\mypara{Research Questions}
In the first experiment, we compare \name with two state-of-the-art baseline tools to evaluate how well they find known binary code vulnerabilities.
In the second experiment, we test how \name interprets the found similar vulnerabilities and how \name assists humans in understanding the reason it considers the query binary vulnerable. 
In the third experiment, we match vulnerabilities in real-world firmware binaries to test how \name work in a real-world application.
In the fourth experiment, we investigate how diverse types of proposed vulnerability signatures (i.e., add, delete, and change) distribute.

\subsection{Performance Metrics}

\mypara{Top-1 Score}
Each vulnerable function was patched after a certain version. 
And all the versions or a range of function versions are vulnerable before that specific version.
Therefore, we select a vulnerable binary function $f$ from binary code $\mathcal{B}$ to test how the tools discover similar vulnerabilities. We construct the vulnerable and patch signature of $f$ from the last pre-patch version and the first post-patch version and store the signature in the database.
To test how well the signature in the database can be matched, we prepare a binary version (denoted by $\mathcal{B}_{v}$) containing the vulnerable version of $f$ (denoted as $fv$), and a patched version binary (denoted by $\mathcal{B}_{p}$) containing the patched version of $f$ (denoted as $fp$) for testing purpose. 

$\mathcal{B}_{V}$ and $\mathcal{B}_{P}$ should differ from the versions that generate the binary signature. $\mathcal{B}_{v}$ and $\mathcal{B}_{p}$ contain many functions, including the vulnerable and patched version of $f$, and other functions. 
For vulnerable function $f$, we inspect each function $fi$ in both $\mathcal{B}_{v}$ and $\mathcal{B}_{p}$ to derive a match score indicating the percentage that $fi$ is similar to $f$'s vulnerable signatures. 
$fv$ in $\mathcal{B}_{v}$ should have the highest score among all other functions; conversely, $fp$ and all other functions in $\mathcal{B}_{p}$ should have low match score. It is reasonable for $fp$ in $\mathcal{B}_{p}$ to have a higher score than other functions in $\mathcal{B}_{p}$ since $fp$ is patched from $f$. 
Nevertheless, $fp$ should be lower than $fv$'s score.
We use the top-1 score to measure the rate of ranking ground truth vulnerable function in the first place.

We provide a simple example of how the top-1 score works in \name. 
Suppose there are ten vulnerable functions, each with a vulnerable and a patched binary version ${B}_{v}$ and ${B}_{p}$. ${B}_{v}$ contains $fv$. ${B}_{p}$ contains $fp$. Both ${B}_{v}$ and ${B}_{p}$ also contain many other functions.
We match the vulnerable signature of $f$ in the database with each function in both ${B}_{v}$ and ${B}_{p}$. 
If vulnerable function $fv$ has the highest score, we rank $fv$ at the top-1 place.  
If 8 out of 10 vulnerable functions rank their testing vulnerable version $fv$ in the top-1 place, then the top-1 score is 0.8. 

\mypara{Mismatch Score}
Merely referring to the top-1 score partly reflects how accurately the tools distinguish ground truth vulnerable functions from other functions. 
However, the top-1 score cannot well demonstrate how the tools consider the non-ground-truth vulnerable function as non-vulnerable. 
A tool that identifies vulnerable functions well with a high top-1 score may not identify non-vulnerable functions as not vulnerable well.
If non-vulnerable functions have extremely close match scores to vulnerable functions, this leads to a high mismatch score.

For instance, some tools may output a similarity score of the ground truth vulnerable function as 0.98, while the score for the ground truth patched function or another random function is 0.97. 
In this case, even though the ground truth function is ranked first, the two scores are too close to reaching the final verdict. 
The ground truth vulnerable function should have a significantly higher score than any other function. 
If any non-ground-truth vulnerable function has a score close to or higher than the ground-truth vulnerable function, the function receives a non-zero \textbf{mismatch score}.

The mismatch score indicates the reliability of the top-1 score.
To keep track of the mismatch score of each vulnerable function, the $\alpha$ parameter is a threshold to activate the mismatch score. 
We consider it a mismatch for any non-vulnerable function with a threshold above $S_{GV}-\alpha$, where $S_{GV}$ denotes the ground-truth vulnerable function score.  
If $S_{GV}$ has an extremely low score (e.g., near zero), any non-vulnerable function having a score close to or above $S_{GV}$ is not considered a mismatch. 
Because the root failure occurs in detecting a vulnerability function rather than non-vulnerable functions, we set $S_{GV} < 0.6$ as an extremely low score.

\subsection{Vulnerability Detection}
Since \name aims to find replicated known vulnerabilities, and the two baseline tools Asm2vec and Palmtree find vulnerabilities based on binary code similarity, we compare \name with those two baseline tools to test their performance on the first objective --- finding replicate vulnerabilities. 
Moreover, the query binary is non-deterministic in real-world scenarios since it could be either a vulnerable or patched version. 
Thus, it is vital to the second objective --- differentiate vulnerable functions from other non-vulnerable functions. 
Therefore, we design this experiment to test these two goals concurrently.

\mypara{Vulnerable Function Detection Accuracy}
\autoref{tab:top-1} lists top-1 scores of \name, Asm2vec, and Palmtree on the seven selected projects. 
Regarding the top-1 score, \name outperforms both baseline methods in six projects and is marginally lower than Palmtree on OpenSSL. 
\name ranks multiple testing vulnerable functions at the top after it extracts accurate vulnerability signatures and matches vulnerable and patched signatures. 
It is because \name matches the fine-grained vulnerability-related instruction (signature) rather than coarsely matches the whole function. 
Since the vulnerability signature tends to be small snippets of instructions, matching the whole function similarity fails to detect such fine-grained information.

\begin{table}[!t]
\centering
\caption{Top-1 scores of seven open-source projects. A higher score indicates a better performance.}
\label{tab:top-1}
\footnotesize
\begin{tabular}{l|c|c|c|c|c|c|c}
\hline
  Project & \rotatebox[origin=c]{270}{Openjpeg} & \rotatebox[origin=c]{270}{FFmpeg} & \rotatebox[origin=c]{270}{Tcpdump} & \rotatebox[origin=c]{270}{Libtiff} & \rotatebox[origin=c]{270}{curl} & \rotatebox[origin=c]{270}{LibPNG} & \rotatebox[origin=c]{270}{OpenSSL}\\
 \hline
 Asm2vec& 0.673& 0.643 &	0.702 & 0.675 & 0.821 & 0.815 & 0.536 \\
 Palmtree& 0.573 & 0.643 & 0.702 & 0.779 & 0.840 & 0.837 & \textbf{0.691}	\\
 \name& \textbf{0.791} & \textbf{0.714} & \textbf{0.786} & \textbf{0.825} & \textbf{0.872} & \textbf{0.859} & 0.673	\\
\hline
\end{tabular}
\end{table}

\mypara{Non-Vulnerable Function Detection Accuracy}
\autoref{tab:mismatch} lists mismatch measurements of \name, Asm2vec, and Palmtree with respect to different $\alpha$ values.  
In the mismatch score perspective, \name achieved the best result with the lowest mismatch score, indicating that \name differentiates the vulnerable version and patched versions with the highest confidence level. 
Conversely, Asm2vec and Palmtree had high mismatch scores, indicating that many decisions between vulnerable and patched versions were made with low confidence. 
Since $\alpha$ denotes the threshold distance to the vulnerable version and $S_{GV} < 0.6$,  we vary $\alpha$ between 0.1 and 0.4 to obtain positive mismatch scores. 
Our evaluation results empirically suggest that $alpha=0.1$ yields the best result.

For FFmpeg, \name achieved mismatch scores as 0 compared with baseline methods' mismatch scores of approximately 1.
For other projects, there are huge contrasts between \name and baseline tools.
\name outperforms two baseline tools because  \name derives vulnerable signatures from the vulnerable and patched versions. 
Another reason is \name matches the fine-grained vulnerability signature rather than the coarse whole function similarity. 
Therefore, subtle vulnerability-related differences are accurately identified, which is superior to whole-function-level similarity matching.

\subsection{Interpretability}


When finding vulnerable functions, a tool's interpretability is as important as its high accuracy. 
In practice, vulnerability-detecting tools assist human experts in making a final verdict. 
Therefore, a good tool should clearly explain why a query result is considered vulnerable.  
Unfortunately, the state-of-the-art baselines fail to provide good interpretation functionality. 
Palmtree outputs only the overall similarity score between the query function and the functions stored in its database. 
In addition to the overall similarity score, Asm2vec lists similar instructions for the query. 
Asm2vec fails to highlight the vulnerability-related instructions; instead, it highlights the whole function as different or similar.

\begin{table}[!t]
\centering
\caption{Mis-match scores of seven open-source projects. A stands for Asm2vec, P for Palmtree, and V for \name. A lower score indicates a better performance.}
\label{tab:mismatch}
\footnotesize
\begin{tabular}{c|c|c|c|c|c|c|c|c}
\hline
 $\alpha$&\rotatebox[origin=c]{270}{Project} & \rotatebox[origin=c]{270}{Openjpeg} & \rotatebox[origin=c]{270}{FFmpeg} & \rotatebox[origin=c]{270}{Tcpdump} & \rotatebox[origin=c]{270}{Libtiff} & \rotatebox[origin=c]{270}{curl} & \rotatebox[origin=c]{270}{LibPNG} & \rotatebox[origin=c]{270}{OpenSSL}\\
 \hline

\multirow{3}{*}{0.1}&A&  0.700 & 0.929 & 0.881 & 0.968 & 0.949 & 0.924 & 0.945\\
&P&	0.909 & 0.821 & 0.905 & 0.955 & 0.750 & 0.946 & 0.936\\
&V&	\textbf{0.091} & \textbf{0.000} & \textbf{0.190} & \textbf{0.162} & \textbf{0.038} & \textbf{0.098} & \textbf{0.118}\\
\hline

\multirow{3}{*}{$0.2$}&A&0.836 & 0.964 & 0.988 & 1.000 & 1.000 & 1.000 & 1.000	\\
&P&0.982 & 1.000 & 0.940 & 1.000 & 0.885 & 1.000 & 0.955	\\
&V&\textbf{0.127} & \textbf{0.000} & \textbf{0.286} & \textbf{0.260} & \textbf{0.103} & \textbf{0.152} & \textbf{0.127}	\\
\hline

\multirow{3}{*}{$0.3$}&A& 0.900 & 1.000 & 1.000 & 1.000 & 1.000 & 1.000 & 1.000\\
&P&0.991 & 1.000 & 0.940 & 1.000 & 0.974 & 1.000 & 0.955\\
&V&	\textbf{0.155} & \textbf{0.000} & \textbf{0.429} & \textbf{0.312} & \textbf{0.147} & \textbf{0.163} & \textbf{0.182}\\
\hline

\multirow{3}{*}{$0.4$}&A&0.962 & 1.000 & 1.000 & 1.000 & 1.000 & 1.000 & 1.000\\
&P&1.000 & 1.000 & 0.976 & 1.000 & 1.000 & 1.000 & 0.964\\
&V&\textbf{0.173} & \textbf{0.000} & \textbf{0.500} & \textbf{0.383} & \textbf{0.224} & \textbf{0.196} & \textbf{0.218}\\
\hline
\end{tabular}
\end{table}

\autoref{fig:sig3} demonstrates an example of \name's interpretability.
This example is a \texttt{many-block-change} vulnerable signature matching selected from CVE-2016-9117. 
The signature (left-hand side) was extracted from the \texttt{imagetopnm} function with versions 2.1.2 and 2.2.0. 
The matched instructions (right-hand side) in the query binary are from version 2.1.1. 
For the selected signature, there are 23 instructions in all structures, and 19 of them are matched. 
The unmatched instructions \texttt{mov rcx, qword ptr [rip + 0xfc246]}, \texttt{mov rcx, qword ptr [rip + 0xfc22b]} on the left-hand side and the instructions \texttt{mov rcx, qword ptr [rip + 0xfc236]}, \texttt{mov rcx, qword ptr [rip + 0xfc21b]} on the right-hand side have different offsets due to structure fields are changed. 
Note that this vulnerable function has multiple signatures, and we omit others for clarity. 
The overall match score combining all signatures exceeds 0.867, indicating \name's high confidence level for the verdict.

\subsection{Real-world Vulnerability Detection}
Since IoT devices' firmware reuse open-source projects, they often contain 1-day vulnerabilities. 
In this experiment, we evaluate how effectively \name detects a real-world 1-day vulnerability in an IoT device's firmware. 
We select four IoT devices' firmware instances (i.e., DCS-3511, DCS-6517, DCS-7517, and DCS-6915) collected in the wild. 
We manually analyze the firmware and prepare 36 ground-truth 1-day vulnerabilities, including 52 vulnerable functions. 
We generate the vulnerability binary code signatures and store them in the database. 
For each vulnerable signature in the database, we detect it against each function $Fi$ in the firmware and assign a matching score for $Fi$. 
If the $Fi$ with the top score is the ground-truth vulnerable function, a vulnerable function is correctly detected. 
\name correctly detects 40 out of 52 (77\%) vulnerable functions. Again, the high accuracy in finding real-world replicate vulnerabilities is due to \name's concentration on the fine-grained vulnerable instructions along with the local control-flow information. We manually analyzed the failed case and found two main failure causes: 1) The binary code contains other function(s) with high similarity to the vulnerable one. 2) The testing binary code contains different structure fields thus at the binary level, the offsets of the structures are different from the signature in the database. For example, \texttt{[esi+0x40]} changed to \texttt{[esi+0x48]} where \texttt{esi} is the memory address of the structure. The same field changed from offset \texttt{0x40} to offset \texttt{0x48} because of adding or deleting other fields in the structure.

\subsection{Statistics of Signature Distributions}

\begin{table}[!t]
\centering
\caption{Vulnerability CWE types of the seven open source projects.}
\label{tab:cwe}
\footnotesize
\begin{tabular}{c|c|c|c|c|c|c|c}
\hline
 & \rotatebox[origin=c]{270}{openssl} & \rotatebox[origin=c]{270}{openjpeg} & \rotatebox[origin=c]{270}{libtiff} & \rotatebox[origin=c]{270}{libpng} & \rotatebox[origin=c]{270}{ffmpeg} & \rotatebox[origin=c]{270}{curl} & \rotatebox[origin=c]{270}{tcpdump}\\
 \hline
NVD-CWE-Other&\textbf{0.15} & 0.01 & 0.04 & \textbf{0.15} & 0.03 & 0.00 & 0.00\\
 CWE-399&\textbf{0.12} & 0.00	&0.01&	0.12	&0.08	&0.00	&0.00\\
CWE-310&\textbf{0.12}	&0.00	&0.00&	0.00&	0.00	&0.00&	0.00\\
CWE-787&0.02	&\textbf{0.15}	&\textbf{0.10}	&0.00&	0.02	&0.00	&0.01\\
CWE-119&0.11	&\textbf{0.33}	&\textbf{0.41}&	\textbf{0.31}&	\textbf{0.37}&	0.00	&\textbf{0.27}\\

CWE-190& 0.01	&\textbf{0.10}&	0.03&	0.02&	0.01&	0.00&	0.01\\
CWE-125&0.03&	0.06&	\textbf{0.13}&	0.02&	0.03	&0.04	&\textbf{0.60}\\
CWE-189&0.04&	0.02&	0.06&	\textbf{0.19}&	\textbf{0.13}	&0.00	&\textbf{0.02}\\
NVD-CWE-noinfo&0.01	&0.01&	0.00	&0.08	&\textbf{0.14}&	0.00&	0.01\\
CWE-126&0.00	&0.00&	0.00&	0.00&	0.00	&\textbf{0.13}&	0.01\\

CWE-122& 0.00&	0.05&	0.00&	0.00	&0.00	&\textbf{0.09}&	0.00\\
CWE-305&0.00&	0.00	&0.00	&0.00&	0.00&	\textbf{0.09}&	0.00\\
\hline
\end{tabular}
\end{table}

In this experiment, we investigate the distribution of the vulnerability according to 1) the Common Weakness Enumeration (CWE) type and 2) our defined three types (i.e., add, delete, change). 
\autoref{tab:cwe} lists the vulnerability distribution according to different CWE types. 
Specifically, we select the three most popular CWE types for each project and concatenate them into the table. 
We observe that \texttt{Improper Restriction of Operations within the Bounds of a Memory Buffer (CWE-119)} is the most common vulnerability type in our experiment (5 in 7 projects). 
Curl contains the most CWE vulnerability types (43 types), while LibPNG contains the least CWE types (11 types).

\autoref{tab:3type} shows the distribution of the four types of vulnerability signatures (i.e., add, delete, one-block-change, and many-block-change). Originally, there were three types (add, delete, change). 
We further split the change type into one-block-change and many-block-change for clarity.
Sig (\#) refers to the number of the signature type in the project. Avg. size refers to the average instruction amount of the specific signature in the project for each CVE.
Generally, \texttt{many-block-change} is the dominant type in all datasets. 
The \texttt{delete} type is the least common type in all datasets. 
The \texttt{add} type contains the most instruction size because the \texttt{add} type involves at least two complete basic blocks to form the signature.
Conversely, the \texttt{delete} type contains the least instruction size because the \texttt{delete} type does not contain control-flow information between multiple blocks that are made up of separate blocks. 
The change types may consist of parent-children structures or separate blocks.

\begin{table}[!t]
\centering
\caption{Statistics of signatures in the 7 open source projects. OBC for one-block-change, and MBC for many-block-change.}
\label{tab:3type}
\footnotesize
\begin{tabular}{cc|c|c|c|c|c|c|c}
\hline
 && \rotatebox[origin=c]{270}{openssl} & \rotatebox[origin=c]{270}{openjpeg} & \rotatebox[origin=c]{270}{libtiff} & \rotatebox[origin=c]{270}{libpng} & \rotatebox[origin=c]{270}{ffmpeg} & \rotatebox[origin=c]{270}{curl} & \rotatebox[origin=c]{270}{tcpdump}\\
 \hline
\multirow{2}{*}{add} &sig (\#) & 120 & 75 & 48 & 10 & 69 & 74 & 248\\
&Avg. size & 50 & 77 & 257 & 46 & 185 & 103 & 61\\
\hline
\multirow{2}{*}{delete} &sig (\#) & 37 & 17 & 9 & 1 & 9 & 20 & 64\\
&Avg.~size & 6 & 3 & 6 & 9 & 10 & 4 & 6\\
\hline
\multirow{2}{*}{OBC}&sig (\#) & 109 & 172 & 61 & 27 & 61 & 85 & 114\\
&Avg.~size & 6 & 11 & 10 & 7 & 14 & 6 & 7\\
\hline
\multirow{2}{*}{MBC}& sig (\#) & 146 & 258 & 91 & 27 & 61 & 109 & 269\\
&Avg.~size & 14 & 23 & 24 & 9 & 20 & 14 & 12\\
\hline
\end{tabular}
\end{table}
\section{Discussion}
\label{sec:dis}

\mypara{Require Source Code} Compared to three state-of-the-art works \cite{vmpbl,binxray,viva}, we require both source code and binary code to extract the signature.
All of the three tools \cite{vmpbl,binxray,viva} claim to only require binary code, but they require all the vulnerability-related versions of binary code, and the binary code must be compiled with the same optimization flag. 
This assumption is strong because one can not guarantee the binary versions (s)he collected from the wild are compiled with the same options.
Therefore, in their actual implementations, they still need the source code to generate different binary codes with the same optimization options from which a signature is extracted.

\mypara{Cross Architecture}  
\name only investigates the vulnerable and patched code on the same architecture.
However, the same source code could be compiled on different hardware architectures (e.g., ARM, x32, PowerPC, etc.) 
How to match cross-architecture vulnerable signatures remains an open research problem. 
Possible solutions include: 1) translating different architectures' instructions into an intermediate language, and 2) extracting vulnerable binary signatures on different architectures. 
However, this issue is beyond this paper's scope.

\mypara{Differences Introduced by Compilation}
An important challenge is mitigating instruction differences introduced by different compiling optimization settings, different compilers, and different compiler versions. 
This paper only considered the project's default optimization options and our testing system's default compiler. 
It is possible to observe the binaries compiled with different optimization levels or compilers in the wild. 
A plausible solution is to utilize symbolic execution to mitigate the impact of different optimization levels as \cite{fiber}. 
However, symbolic execution is time-consuming to execute. 
Another possible solution without changing our current methodology is to increase our training data. The training data refers to the binaries we extract signatures from.
Since we only extract vulnerability signatures from vulnerable and patched versions compiled by their default optimization level and the default compiler, the current training data are limited. 
To detect cross-optimization-level or cross-compiler signatures, a possible solution is to compile the project using multiple optimization levels or compilers and extract their corresponding signatures.

Patch and vulnerability detection genres of work directly extract assembly instructions and form signatures. 
The state-of-the-art whole-function similarity matching adopts many data-driven methods. 
Asm2vec \cite{asm2vec} and Palmtree \cite{Palmtree} convert the assembly instructions into vectors to mitigate subtle assembly differences introduced by compilations to some extent. 
Data-driven methods usually take less time than other methods. 
Merging these two methods by generating vectorized fine-grained signatures detects fine-grained signatures and mitigates assembly differences with less time and cost. 
Graph attributes-based vectors are generated in \cite{gemini,VULSEEKER}. 
Therefore, it is possible to extend \name by incorporating fine-grained graph-based embeddings as the signature.

\section{Related Work}
\label{sec:related}
We present the related work from the following threefold since they are closely related to this work: 1) code similarity detection, 2) patch analysis, and 3) vulnerability detection.

\subsection{Code Similarity Detection}

\subsubsection{Binary-code-level similarity detection} 

Binary-code-level similarity works are categorized in two directions according to their methods.

\mypara{Learning-based methods} Binary code instructions are encoded into an embedding to compare the similarity. 
Gemini \cite{gemini}, Vulseeker \cite{VULSEEKER}, and Genius \cite{genius} use graph feature embeddings to determine vector similarity. 
Safe \cite{safe}, InnerEye \cite{innereye}, $\alpha$Diff \cite{aDiff}, Kam1n0 \cite{Kam1n0}, and Asm2Vec \cite{asm2vec} learn the instructions' embeddings and generate block embeddings or function embeddings.

\mypara{Program-analysis based methods}
Instructions or blocks are regarded as sequences in Binsequence~\cite{binsequence} and Tracy~\cite{tracy} using sequences-alignment methods to compare the similarity.
Similarly, SIGMA~\cite{SIGMA}, FOSSIL~\cite{fossil}, and Beagle~\cite{BEAGLE} rely on the instruction semantic categorizations like data transfer, logic, or arithmetic. 
Bingo~\cite{bingo} and IMF-SIM~\cite{IMF-SIM} use input-output relations to measure binary code similarity. 
Expose~\cite{Expose}, Binhash~\cite{binhash}, Binhunt~\cite{binhunt}, CoP~\cite{COP}, ESH~\cite{esh}, GITZ~\cite{GITZ}, and XMATCH~\cite{xmatch} symbolically execute the binary code before the similarity comparison based on symbolic formulas.

\mypara{Limitations} 
However, similarity-based methods match the whole function similarity. 
Vulnerable instructions only involve several lines of code in the function. 
Therefore, the similarity-based method can filter similar functions but cannot distinguish whether the function is vulnerable.

\subsection{Patch Identification and Analysis}
FIBER \cite{fiber} detects patch existence in Linux kernel binaries based on symbolic execution. 
Using symbolic execution and memory status, PDiff \cite{pdiff} detects Linux kernel binaries' patch existence when binaries are different due to patch customization, different build configuration, and other reasons.
Spain \cite{spain} uses binary-level semantic information to identify the patch before summarizing patch and vulnerability patterns. 
Patchscope \cite{patchscope} identifies patch existence based on memory-object-centric methods and dynamic execution.

\mypara{Limitations} 
This category of prior work assumes that the function names are provided or that some similar candidate functions have already been selected by the code-similarity-based method. 
Moreover, they focus on patch detection rather than vulnerability detection. 
The lack of a patch does not necessarily imply that the function is vulnerable.

\subsection{Vulnerability Detection}
VMPBL \cite{vmpbl} builds a database storing vulnerable and patched functions to distinguish the pre-patch and post-patched functions. 
VIVA \cite{viva} collects binary with versions before and after the patch and directly diff the pre-patch and post-patch functions to retrieve binary-level vulnerability signatures. 
VIVA further detects vulnerability existence based on pre-filtering and instruction clustering. 
BINXRAY \cite{binxray} requires pre-patch and post-patch version binaries to analyze the vulnerability-related instructions in both versions before storing instructions in the database as vulnerability and patch signatures. 
BINXRAY checks the vulnerability's existence in a query function based on its closest signature version.

\mypara{Limitations} 
This genre of work is most similar to our methods. 
However, they assume all the different binary codes between versions are related to the vulnerability, this often introduces many vulnerability-irrelevant instructions into signatures.
\section{Conclusion}
\label{sec:ccln}

In this paper, we proposed a novel approach called \name to extract and match binary-level vulnerability-related signatures. 
\name consists of four steps: 1) data preparation, 2) locating signature instruction, 3) constructing context-aware binary-level signatures, and 4) signature matching. 
Compared to previous work, \name accurately locates vulnerability-related instructions and detects vulnerability within functions. 
Through our empirical studies, \name outperformed two state-of-the-art similarity-based vulnerability detection tools --- Asm2vec \cite{asm2vec} and Palmtree \cite{Palmtree}.
Specifically, \name achieved the most accurate results on six out of seven projects with the least ambiguities while providing reasons for vulnerable functions. 
Hence, \name effectively facilitates human understanding of its decision process during vulnerability detection. 
Our experiment on real-world firmware vulnerability detection indicates \name is practical to find vulnerabilities in real-world scenarios.
Our analysis of vulnerability distributions confirmed that \name is a versatile detector with good potential for future extension. 


\bibliographystyle{IEEEtran}
\bibliography{ref}

\newpage
\appendix
\section{Appendix}
\subsection{Considerations of Header Files} 
In 
Section III B:Locating Signature Instructions and Challenges, when we locate the vulnerable binary code from source code, in some cases, certain CVEs include source code changes in the corresponding header (.h) files. 
However, such CVEs are less frequent in number (i.e., 26 CVEs in total out of nearly 1000 CVEs). 
After manually analyzing the changes in the header (.h) files due to the addition of patching codes, we discovered three types of changes, including \textbf{1) Change in MACRO values}, \textbf{2) change in structure member variables} (i.e., changing, adding, or deleting structure member variables, and \textbf{3) change in the function definition}.

\mypara{Change of MACRO value.}
Diffing the source code versions cannot detect MACRO value changes in the .h file (e.g., \texttt{\#define HAVE\_DIRENT\_H 1} to \texttt{\#define HAVE\_DIRENT\_H 0}).
However, we omit this concern due to the tiny number of the MACRO changing related CVEs (i.e., 8 out of nearly 1000 CVEs). 

\begin{figure}[!h]
\centering
\includegraphics[width=0.47\textwidth]{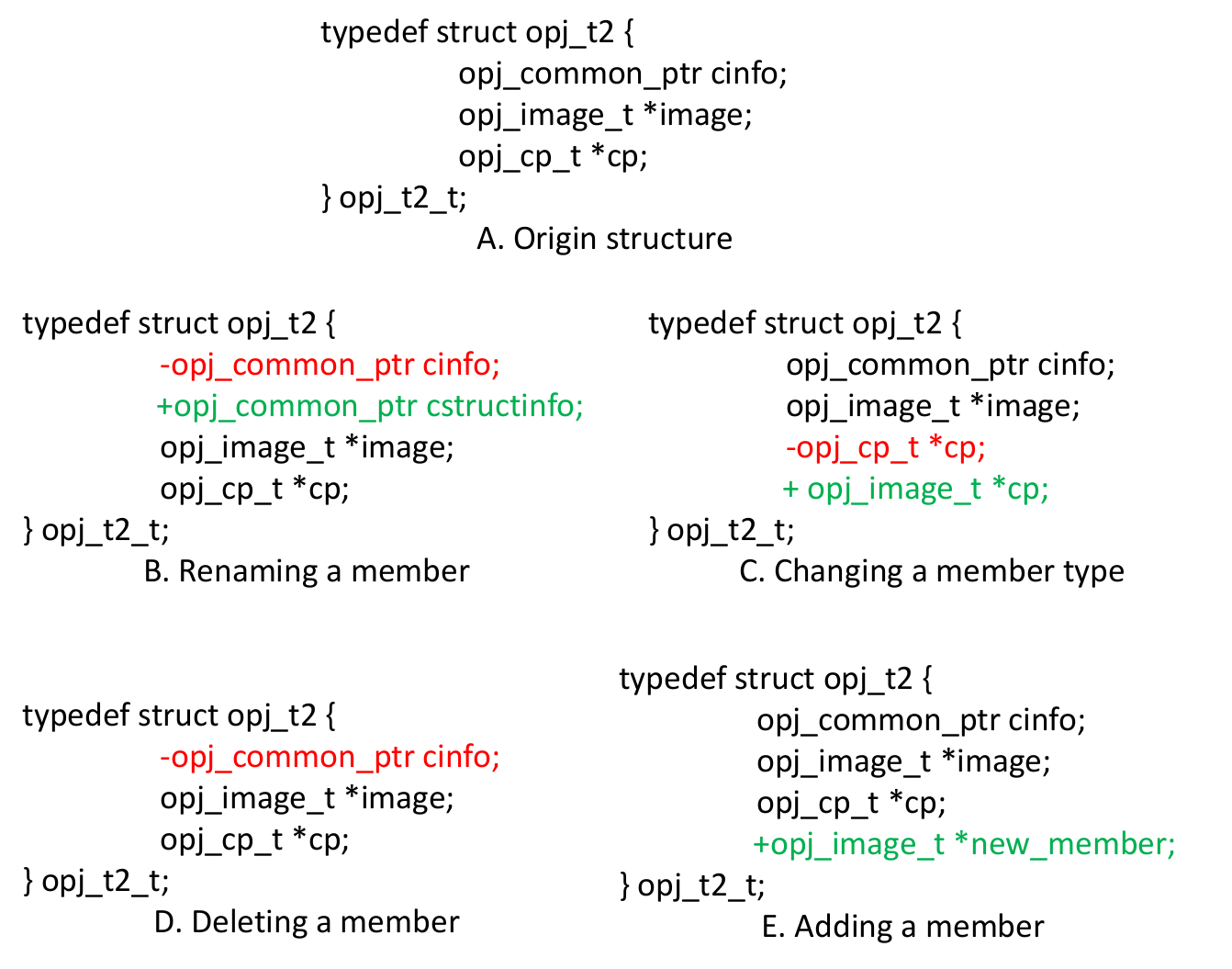}
\caption{Examples of structure member changes. A is the original structure in the vulnerable version. B to E are the examples in the patched version. Red fonts mean the instructions to be deleted, and green font means the new instructions to be added in the patched version. The example is from openjpeg version 1.5.0.}
\label{fig:struct_change}
\end{figure}

\mypara{Change of structure member.}
Structure members can be modified in a few ways --- changing, adding, or deleting. 
\autoref{fig:struct_change} demonstrates some examples for each type.
Specifically, changing structure members include renaming the member and changing the type of existing member. \autoref{fig:struct_change} B and C shows two examples of renaming member and changing member type, respectively.
The source code with red text font represents the code to be deleted, and the green font means the instructions to be added in the patched version. 
If a structure member is deleted, the .c files source codes mentioning the member must also be deleted. 
Those source code line changes can be detected by diffing the .c files. 
For the adding or changing structure members cases, it is difficult to detect the changes by diffing the .c source codes. 
For example, assume that we observe a structure member is added, as shown in \autoref{fig:struct_change} E, one may think that the added structure member \texttt{new\_member} may not have a corresponding update in the .c file referencing it. 
To identify this case's frequency, we manually inspected all 26 CVEs.
We found that when members are added or changed in the structure, there must be corresponding new source code lines updated in the .c files referencing them.
Therefore, in approximately 1000 vulnerability functions, we found in 100\% cases, the changed or added members have corresponding references in the .c files.
A possible explanation is that the newly added or changed structure members are specifically designed to be used in the .c files to avoid vulnerabilities. \autoref{fig:struct_change} D and E show two examples of deleting and adding a structure member, respectively.
In the approximate 1000 vulnerabilities, we observe that all the structure member changing, adding, or deleting can be detected with the diff tool.

\mypara{Change of function definition.}
Changing of function definition refers to changing function calling parameters (e.g., change function definition \texttt{static void j2k\_write\_sot(opj\_j2k *j2k)} to \texttt{static void j2k\_write\_sot(opj\_j2k *j2k, int lenp)}).
This category of change can be reflected in the source code. 
Function calling parameter changes can be detected in the .c files referencing that function because the source code must be updated to handle different parameters. 
Since we are extracting vulnerable function signatures between the vulnerable and patched versions, newly added functions and deleted functions are out of scope because they either only exists in the vulnerable version or only in the patched version.

\begin{IEEEbiography}[{\includegraphics[width=1in,height=1.25in,clip,keepaspectratio]{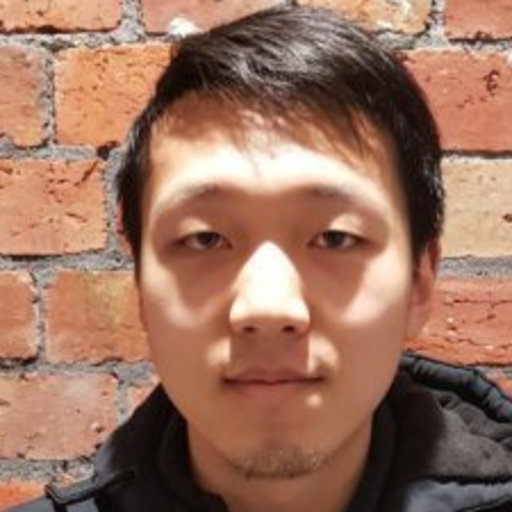}}]{Zian Liu}
received the bachelor of information
technology degree from Deakin University Australia,
in 2018. He is currently working towards the PhD
degree at the Swinburne University of Technology.
His research interests include binary code analysis,
especially in vulnerability detection.
\end{IEEEbiography}

\begin{IEEEbiography}
[{\includegraphics[width=1in,height=1.25in,clip,keepaspectratio]{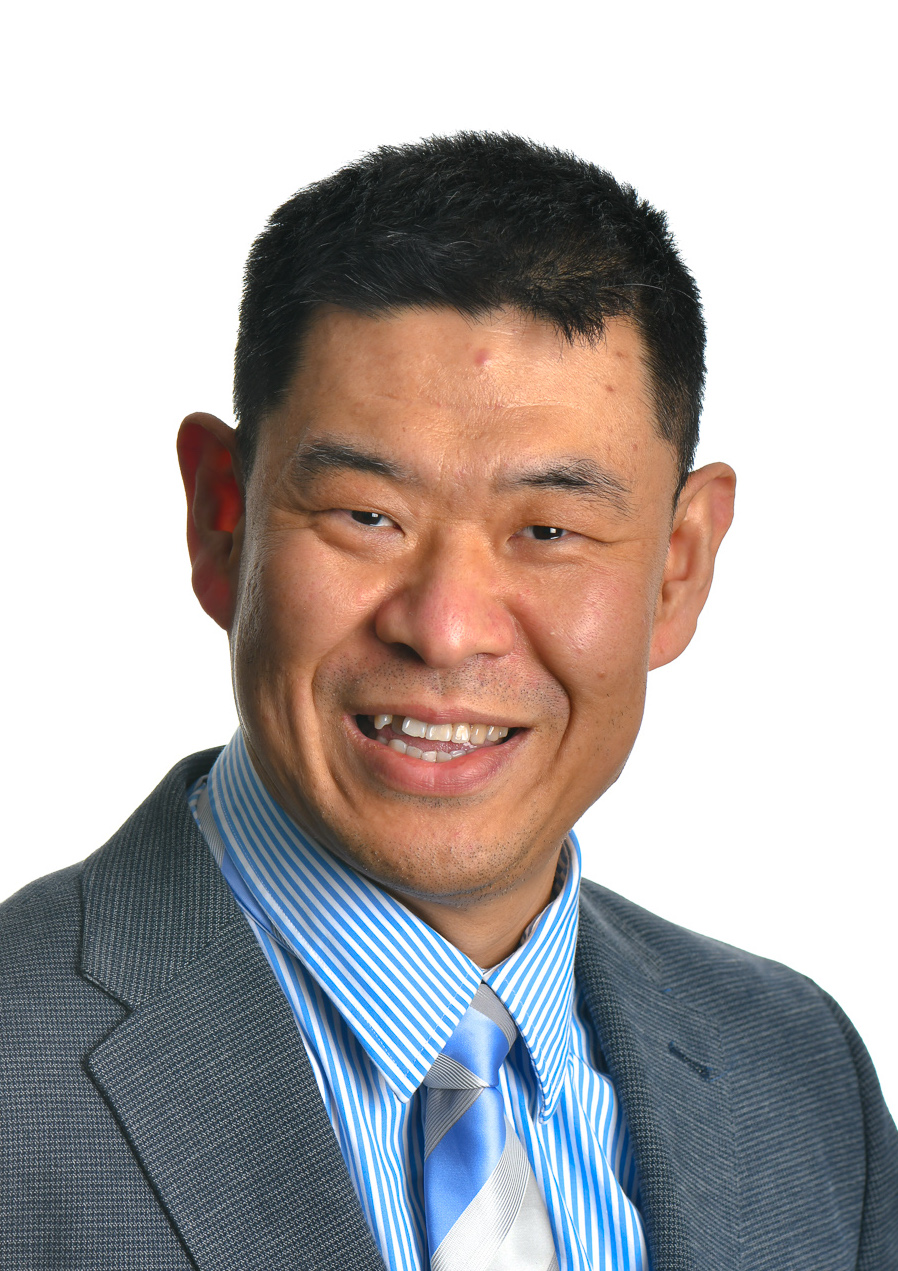}}]{Lei Pan}
received the Ph.D.~degree in computer forensics from Deakin University, Australia, in 2008. He is currently a Senior Lecturer with the Centre for Cyber Resilience and Trust (CREST), School of Information Technology, Deakin University. He leads the research theme `Securing Data and Infrastructure' at CREST. His research interests cover broad topics in cybersecurity and privacy. He has authored 100 research papers in refereed international journals and conferences, such as IEEE Transactions on Information Forensics and Security, IEEE Transactions on Dependable and Security Computing, IEEE Transactions on Industrial Informatics, and many more.
\end{IEEEbiography}

\begin{IEEEbiography}
[{\includegraphics[width=1in,height=1.25in,clip,keepaspectratio]{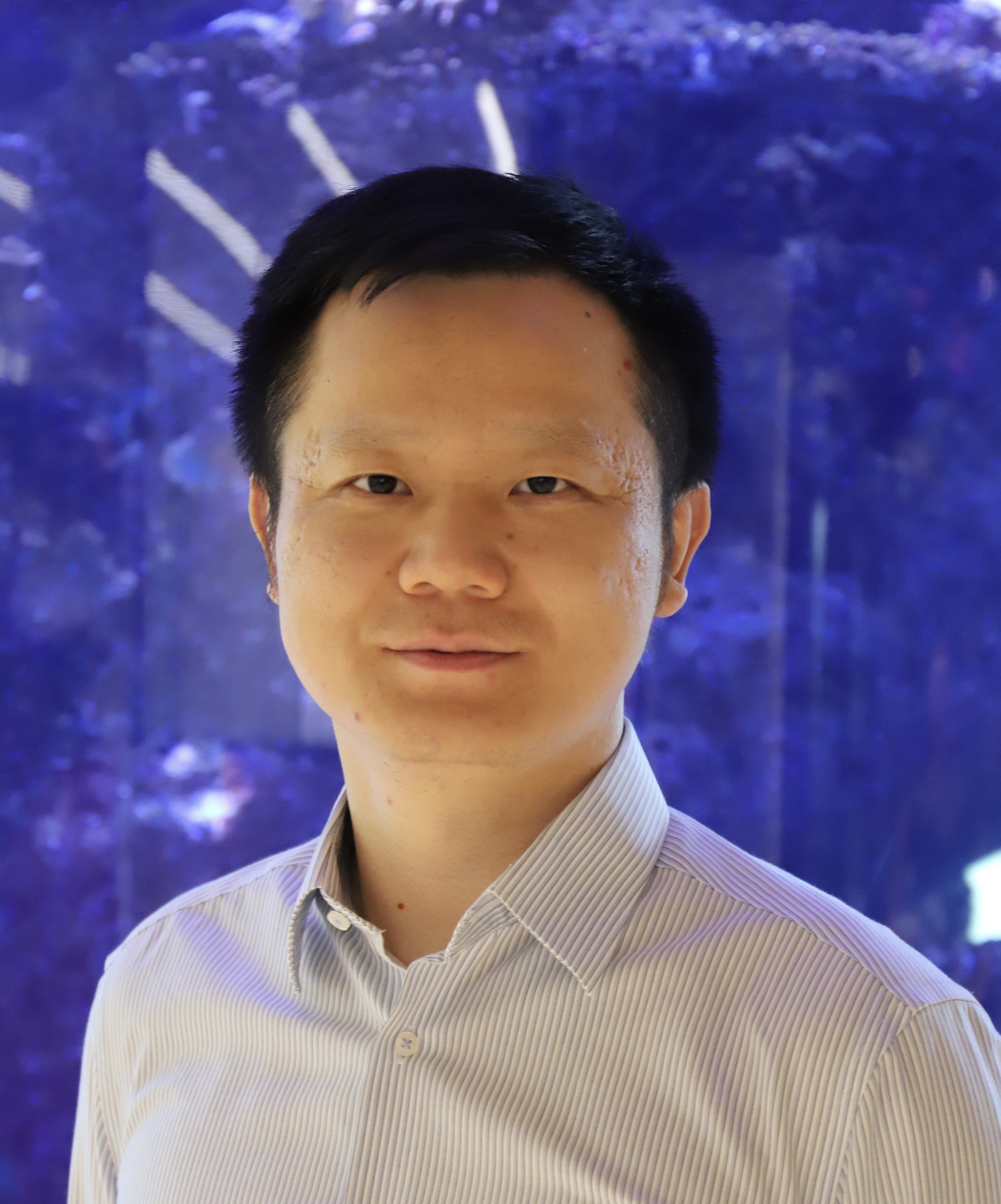}}]{Chao Chen} received his PhD degree in Information Technology from Deakin University in 2017. He is focusing on using AI and advanced analytics to solve real-world problems, such as network traffic analysis for abnormal behaviour, social spam detection, insider threat detection, and software vulnerability. He is also conducting research on responsible AI, such as the transparency and trustworthy of AI applications in enterprises. He has published more than 40 research papers in refereed international journals and conferences (with 16 Q1 journals), such as ACM Computing Surveys (CSUR), IEEE Transactions on Information Forensics and Security (TIFS), Privacy Enhancing Technologies Symposium (PETS) and ACM Asia Conference on Computer \& Communications Security (ASIACCS). One of his papers was the featured article of that issue (IT Professional Mar.-Apr. 2016).
\end{IEEEbiography}

\begin{IEEEbiography}
[{\includegraphics[width=1in,height=1.25in,clip,keepaspectratio]{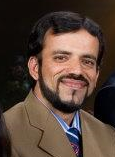}}]
{Muhammad Ejaz Ahmed} is a Senior Research Scientist at Data61, CSIRO, Australia's national science agency. His research interests include malware (ransomware) detection, threat hunting, digital forensics, program analysis, natural language processing, and machine learning. He received the B.Sc. degree in computer sciences from Peshawar University, and the M.S. degree from National University of Sciences and Technology (NUST), Pakistan in 2006 and 2011, respectively. He completed the Ph.D. degree from Electronics and Radio Engineering Department at Kyung Hee University of South Korea in February 2014. Prior to that, he was postdoc with POSTECH South Korea from June 2014 to May 2015. He was with Sungkyunkwan University of South Korea as a research professor from June 2015 to May 2018.
\end{IEEEbiography}

\begin{IEEEbiography}
[{\includegraphics[width=1in,height=1.25in,clip,keepaspectratio]{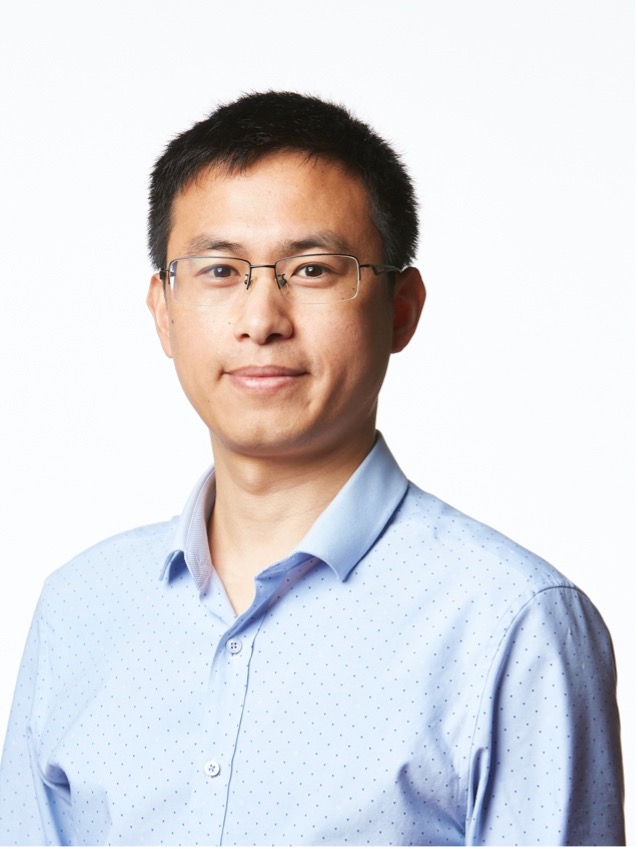}}]{Shigang Liu}
(M'15) received his PhD in Computer Science from Deakin University, Australia in 2017. He is currently a research fellow at the School of Science, Computing and Engineering Technologies at Swinburne University of Technology. His research primarily focuses on data-driven software security, network security, applied machine learning, and fuzzy information processing. In 2019, his research won first place in the World Change Maker Prize at the Swinburne Research Conference. He has served as Program Chair for various international conferences such as CSS2017/2020/2022, NSS2020/2022, ML4CS2019, SocialSec2022, IEEE Blockchain2022, and so on.
\end{IEEEbiography}

\begin{IEEEbiography}
[{\includegraphics[width=1in,height=1.25in,clip,keepaspectratio]{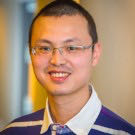}}]{Jun Zhang}
(M’12-SM’18) received the Ph.D. degree in Computer Science from the University of Wollongong, NSW, Australia, in 2011. He is currently a full Professor and the Director of the cybersecurity lab, Swinburne University of Technology, Australia. He was recognized in The Australian’s top researchers special edition publication as the leading researcher in the field of Computer Security \& Cryptography in 2020. He leaded Swinburne cybersecurity research and produced excellent outcome including many high impact research papers and multi-million-dollar research projects. Swinburne was named in The Australian's 2021 Research magazine, the top research institution in the field of Computer Security \& Cryptography. He has served as a steering committee member of the P-TECH program at Melbourne since 2019, which the Australian Government invested in, promoting STEM education. He devotes himself to communication and community engagement, boosting the awareness of cybersecurity.
\end{IEEEbiography}

\begin{IEEEbiography}
[{\includegraphics[width=1in,height=1.25in,clip,keepaspectratio]{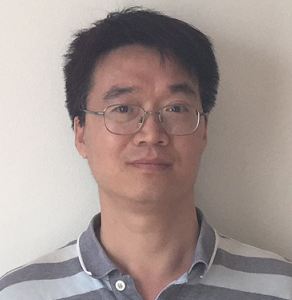}}]{Dongxi Liu} is a Principal Research Scientist in CSIRO’s Data61, joined CSIRO since 2008. His research interests include applied cryptography, post-quantum cryptography, and distributed system security. His work aims to design and build secure systems that are scalable, simple to use, with high trust to security and resilient to attacks.
\end{IEEEbiography}

\end{document}